\documentstyle[11pt,aaspp4]{article}
\def \hb {H$\beta$}

\def \oiii  {{\rm [O~III]}}

\def\wave#1{$\lambda${#1}}
\def\waves#1{$\lambda\lambda${#1}}
\def\etal{{\rm et al.}}

\def\arcsecpoint{\hbox to 1pt{}\rlap{\arcsec}.\hbox to 2pt{}}
\def\ltsima{$\; \buildrel < \over \sim \;$}
\def\simlt{\lower.5ex\hbox{\ltsima}}
\def\gtsima{$\; \buildrel > \over \sim \;$}
\def\simgt{\lower.5ex\hbox{\gtsima}}
\def\samename{\vrule height0.4pt depth0.0pt width1.0in \thinspace.}
\def   \flam    {$F_\lambda$}
\def   \pf      {$P\times F_\lambda$}

\def   \kms     {km s$^{-1}$}
\def   \p       {$P$}

\def   \pa      {$\theta$}

\received{18 August 1997}
\accepted{21 December 1997}
\lefthead{Tran et al.}
\righthead{Scattered Radiation in Radio Galaxies}

\slugcomment{{\it Astrophysical Journal}, accepted 21 December 1997}

\begin{document}

\title{Scattered Radiation from Obscured Quasars in Distant Radio Galaxies}
\author{Hien D. Tran\altaffilmark{1}, Marshall H. Cohen\altaffilmark{2},
Patrick M. Ogle\altaffilmark{2}, 
Robert W. Goodrich\altaffilmark{3}, 
Sperello di Serego Alighieri\altaffilmark{4} }
%\altaffiltext{1}{UCO/Lick Observatory, University of California, Santa Cruz,
%CA 95064; tran@ucolick.org.}
\altaffiltext{1}{Institute of Geophysics and Planetary
Physics, Lawrence Livermore National Laboratory, 7000 East Avenue, P.O. Box
808, L-413, Livermore, CA 94550; htran@igpp.llnl.gov.} 
\altaffiltext{2}{Palomar Observatory, California Institute of Technology, 
Pasadena, CA 91125; mhc@astro.caltech.edu, pmo@astro.caltech.edu.}
%\altaffiltext{4}{Space Telescope Science Institute, 3700 San Martin Drive, 
%Baltimore, MD 21218; martel@stsci.edu.}
\altaffiltext{3}{CARA/Keck Observatory, 65-1120 Mamalahoa Highway, Kamuela, HI 
96743; goodrich@keck.hawaii.edu.}
\altaffiltext{4}{Osservatorio Astrofisico di Arcetri, Firenze, Italy; 
sperello@arcetri.astro.it.}
%\altaffiltext{7}{Space Telescope--European Coordinating Facility, European
%Southern Observatory, Karl-Schwarzschild Str. 2, D-85748, Garching bei Munchen,
%Germany; rfosbury@eso.org.}

\begin{abstract}

We present optical spectropolarimetric and imaging polarimetric 
observations of four high-redshift radio galaxies (HzRGs) obtained with 
the Low Resolution Imaging Spectrometer (LRIS) of the 10-m Keck I telescope.
A broad Mg II \wave 2800 emission line is detected in the total and polarized
flux spectra of 3C 265 and 3C 277.2. The fractional polarization is high, and 
both it and the position angle are constant with wavelength after accounting for
dilution by unpolarized starlight of the host galaxy, which can contribute
substantially. An extended unpolarized continuum similar
to that observed in other AGNs is also detected. Imaging polarimetry
reveals a rough double-fan morphology of the polarized light coincident 
with the extended aligned emission regions, with
the position angle essentially perpendicular to the radial structure of the
extended UV/optical emission, and with the 
degree of polarization increasing with radius away from the nucleus. 
The radio jets lie inside the extended emission regions and, like every
radius, are roughly perpendicular to the polarization PA.
These results strengthen the view that powerful radio galaxies would be
called quasars if viewed from a proper direction.
Based on the polarimetric data
presented in this paper and in previous studies, 
scattering of radiation from an obscured quasar source appears to be
the preferred interpretation over jet-induced star formation for 
explaining the alignment effect in HzRGs. Both electrons and dust 
can play a major role in the scattering process.
However, the lack of strong direct evidence for either case, and 
our ignorance of the properties and distribution of 
the scatterers in these galaxies make it very difficult to discriminate
between the two. Our data reveal a chance alignment of 3C 343.1
with a foreground galaxy, which dominates the observed optical flux from
the system. 

\end{abstract}

\keywords{galaxies: active --- galaxies: individual (3C 265, 3C 277.2, 3C
324, 3C 343.1) --- polarization --- radio sources: galaxies}

\section{Introduction}

Distant radio galaxies ($z$ \simgt 1.0) offer a powerful laboratory for 
probing cosmology and galactic evolution, because they are extragalactic 
objects that are spatially extended and visible at large look-back times. 
They are also important because they appear to be related to 
quasars (based on the radio galaxy/quasar unification model; 
e.g., Barthel 1989), 
thus providing important insights into understanding the nature of
this most powerful class of objects in the universe. 
The discovery that many distant radio galaxies
show extended emission-line and UV-continuum emitting regions that are often 
aligned with the radio structure axis [the ``alignment effect", 
see McCarthy (1993) for a review] means that this phenomenon must be
understood before these galaxies can be used as probes of cosmology and 
galactic evolution. Several suggestions have been put forward
to account for this effect.
They include star formation 
resulting from the interaction of the radio jets with the interstellar medium 
(De Young 1981, 1989; Chambers, Miley, \& Joyce 1988; Rees 1989) 
and alternatively, nuclear light scattered by dust or electrons 
(e.g., Tadhunter \etal~1992; 
Cimatti \etal~1993; di Serego Alighieri, Cimatti, \& Fosbury 1994). 
Another, poorly investigated possibility is that the extended emission
represents the illumination pattern of a Doppler-beamed continuum, analogous
to that seen directly in blazars.
Other mechanisms are also possible (see, e.g., Daly 1992) but currently 
these are the main contentions. 
Polarimetric observations showing that the extended 
emission is highly polarized, with the polarization position angle (PA)
perpendicular to the radio structure axis 
(e.g., Jannuzi \& Elston 1991; Tadhunter \etal~1992; Cimatti \etal~1993; 
di Serego Alighieri \etal~1994) 
offer strong support for the scattering pictures, but do not totally exclude 
star formation. For example, a large amount of dust associated with star 
forming regions along the jets could also give rise to scattering of radiation
from the central nucleus in the jet-induced star formation scenario. 
There have also been other arguments for this picture 
(e.g., Best, Longair, \& R\"ottgering 1996, 1997). 

If the scattering picture is correct, we expect the extended light to be 
highly polarized, and when combined with the radio galaxy/quasar hypothesis, 
we also expect to see a quasar spectrum in polarized light. Just such
a picture indeed has been shown to be the case in some of the 
low-redshift radio galaxies (i.e., 3C 234, Antonucci 1984; 
Tran, Cohen, \& Goodrich 1995; 3C 321, Young \etal~1996; Cygnus A, Ogle
\etal~1997). Tran \etal~also showed that the 
radio galaxy 3C 234 would be indistinguishable from a quasar if 
it were not for the high amount of obscuration in the
nucleus, and a similar result was found for 3C 109 (Goodrich \& Cohen 1992). 
A similar unification model based on obscuration/orientation 
effects has been well-established for Seyfert 1 and 2 galaxies 
(e.g., Antonucci \& Miller (1985); Miller \& Goodrich (1990); Tran 1995a, b, c).
Recent observational evidence is mounting
that the same picture may also apply to many high-redshift radio galaxies 
(HzRGs) (di Serego Alighieri \etal~1994; Dey \& Spinrad 1995; 
Cimatti \etal~1996, 1997; Cohen \etal~1996; Dey \etal~1996) based on 
the detection of the broad Mg II \wave 2800 line in either the total flux 
spectra, or polarized flux spectra.
A convincingly detected broad Mg II line in the polarized
flux spectrum would demonstrate that the quasar scattering hypothesis and
the obscuration/reflection picture are at least partly responsible for
the alignment effect and that radio galaxies may be quasars viewed from an
equatorial direction. 
Di Serego Alighieri \etal~(1994) and Cimatti \etal~(1996, 1997) have 
presented evidence for broad Mg II in the polarized-flux spectra of 
some of these galaxies, but these results are based on fairly low 
signal-to-noise ratio (S/N) data, and further study is needed. 

Alternatively, if the extended continuum represents scattering of a Doppler
beamed source, the polarized light should consist predominantly of continuum, 
with little or no emission lines. We expect significant Doppler beaming in 
all quasars, as quasars viewed from polar directions are thought to appear 
as blazars. A hybrid of the last two scenarios is likely, in which the 
scatterers see a normal quasar spectrum with some contribution, especially 
in the polar directions, by a beamed continuum. Ultimately one might hope 
to constrain beaming models via spatially resolved spectropolarimetry. 

While the mechanism responsible for the polarized light in 
these HzRGs is generally thought to be scattering, the nature of the 
scatterers continues to be a source of debate. At issue is whether 
scattering by electrons or dust dominates, although it is recognized 
that {\it both} could be present in producing the observed polarization. 
The nature of the scattering agents could have some important implications 
on the alignment effect and nature of these galaxies.
For example, if it can be conclusively shown that electron scattering
is primarily responsible for the extended polarized light, it would provide 
an argument against jet-induced star formation as the main cause 
of the alignment effect, since this picture predicts that dust grains that 
are abundant within the star forming regions in the jet streams would be 
the principal medium for scattering the light. 

In this paper, we investigate a number of HzRGs using the Keck I 10-m 
telescope to address various issues relating to the alignment effect,
the nature of the scatterers, and the radio galaxy/quasar unification 
hypothesis. In particular, we look for the signature of broad Mg II 
in polarized flux, and map the polarization structure of the extended 
emission in order to investigate the central 
nucleus and the geometry of the scattering process. 
We discuss the implications of these observations on the above issues. 
Our sample of objects comprises those with previously detected polarizations:
3C 265, 3C 277.2 and 3C 343.1, which are slightly brighter and have relatively 
smaller redshifts 
($z \sim$ 0.8) than those ($z \sim$ 1.4) studied by Dey \etal~(1996), and
Cimatti \etal~(1996, 1997) (3C 13, 3C 256, 3C 324, 3C 356), 
and our study complements the previous work.
We also use imaging polarimetry to study 3C 324. 

Section 2 describes the observations. In \S~3 we present our data analysis
and results of the spectro- and imaging polarimetry. The discussion
and interpretation of these results are presented in \S~4, and  
a summary and conclusions in \S~5.

\section{Observations}

Our observations were carried out at the W. M. Keck Observatory using
the 10-m telescope with the polarimeter (Cohen \etal~1997) installed in the 
Low Resolution Imaging Spectrometer (LRIS, Oke \etal~1995). 
A 1\arcsec~wide, long slit was centered on the nucleus of the galaxy 
and oriented along the UV extensions. 
A 300 grooves mm$^{-1}$ grating provided a dispersion of
2.5 \AA~pixel$^{-1}$ and a resolution of $\sim$ 10 \AA~(FWHM). 
The spectropolarimetric observations were performed
following standard procedures of rotating the half waveplate to four
position angles, and calibrating the instruments with null and standard 
polarization stars (see Cohen \etal~1997 for details). 
No second-order blocking filter was used in these observations, but
we observed the flux-standard stars both with and without a GG495 
filter, which blocks all light with $\lambda < 4950$ \AA, 
to correct for any second-order light. The residual second-order effects 
are small and apply only to the spectral regions $\simgt$ 7500 \AA. 
They do not affect the conclusions drawn in this paper.

Table 1 shows the log of the observations. The choice of objects was 
based on their relatively high apparent brightness ($m_V < 22$), 
high polarization from surveys 
published in the literature (Cimatti \etal~1993), or display of 
dramatic alignment effect. To improve S/N, multiple observations of the same 
object were carried out at different epochs, and the individual results 
coadded and presented here as the final form. 

In addition to spectropolarimetry, we also obtained imaging 
polarimetric observations for three galaxies: 3C 265, 2C 277.2 and 3C 324. 
The imaging polarimetry used procedures identical 
to those employed in spectropolarimetry, except they were made through 
either a standard $B$ or $V$ filter, no slit, and with a mirror replacing the 
grating. Data reductions were carried out using the imaging processing 
software VISTA as described by Tran \etal~(1995) and Cohen \etal~(1997). 

\section{Results and Analysis}

\subsection{3C 265}

Of all the HzRGs that we observed, 3C 265 ($z$ = 0.811) is the brightest 
($m_V = 20.9$) and has by far the best S/N data. 
We thus focus our discussion and analysis on 
this object. The total flux spectrum of 3C 265 in the spectral region around 
Mg II has been discussed by Dey \& Spinrad (1996). 
The nuclear spectrum, shown in Figure 1, is typical of a narrow line 
radio galaxy (NLRG), displaying many emission lines with a wide 
range in strengths, ionizations and atomic species
[compare, for example, to the spectrum of 3C 234 (Tran \etal~1995)]. 
In Table 2 we list the measured emission line ratios (relative to H$\beta$)
and the observed equivalent widths (EWs) from the nuclear spectra of 3C 265, 
along with those of 3C 277.2 and 3C 343.1. 
Except for 3C 343.1, these measurements were made from the observed 
spectrum without removing the host starlight. 
Detailed analysis of these spectrophotometric measurements in relation to 
dust and electron scattering and photoionization models will be the subject 
of a later paper. 

In Table 2 we identify several faint UV lines that are rarely observed in 
a HzRG. These include [Mg VII] \waves 2509, 2629
(equivalents of [O III] \waves 4959, 5007 in the same isoelectronic sequence),
[Mg V] \waves 2783, 2928 (equivalents of [O I] \waves 6300, 6364),
as well as other lines due to permitted He II and O III
resonance-fluorescence.
We identify a weak, broad component due to the resonance-fluorescence line
O III \wave 3444 at the base of [Ne V] \wave 3426. 
Such a broad component was also noted by Tran \etal~(1995) in the 
polarized flux spectrum of 3C 234. 

Dey \& Spinrad (1996) identified the emission line blueward of Mg II
\waves 2796, 2805 as [Mg VII] \wave 2786.
It is more likely, however, that this line is actually [Mg V] \wave 2783
(Osterbrock 1963, 1997).
Since the two [Mg VII] lines \waves 2509, 2629 have been identified
in our spectrum and they are analogous to the strong [O III] \waves 4959, 5007
ubiquitous in nebular spectra, it is unlikely that there is a third [Mg VII]
line stronger than these two.
Furthermore, the [Mg V] \wave 2783 is equivalent to [O I] \wave 6300 in the
same isoelectronic sequence, and if its identification is correct, we would
expect to see the analog of [O I] \wave 6364 in [Mg V] at \wave 2928 with
about 1/3 intensity ratio. Such a line is indeed present with about the
correct intensity ratio (see Table 2). Thus, there is little doubt that
the emission line at 2783 \AA~is correctly identified with [Mg V].
Its identification shows that the very wide range of ionization in 3C 265 is
continuous.
The feature just to the red of the Mg II doublet has a stubby profile,
suggesting that it is not a single line but probably is a doublet.
We identify it as the blend He I \wave 2830 + O III \wave 2836.

In Figure 1 broad wings in Mg II can clearly be seen in the total flux 
spectrum. 
A least-squares fit of multiple Gaussian profiles to the
Mg II emission-line complex is shown in Figure 2. The 
FWHM of the broad Mg II in total flux is 11,000 $\pm$ 700 \kms, 
similar to but somewhat broader than that measured by Dey \& Spinrad (1996). 
The spectropolarimetry of the nuclear region of 3C 265 is
shown in Figure 3. The observed fractional polarization $P$ 
(Fig. 3b) is high, rising smoothly from 6\% in the red 
(rest-frame blue) to $\sim$ 12\% in the blue (rest-frame UV).
Broad Mg II emission is clearly present in the polarized 
flux spectrum \pf~(Fig. 3d; see also Cohen \etal~1996), 
demonstrating that the broad line photons seen in the total flux spectrum 
are polarized, presumably by scattering of the light from an obscured nucleus. 
Note that the broad O III \wave 3444 line also appears to be present in 
the \pf~spectrum of 3C 265 (Fig. 3d), similar to that observed in 3C 234
(Tran \etal~1995).
By contrast, the narrow lines are essentially unpolarized, as
evidenced by the sharp drop in $P$ at these lines and their disappearance 
in the polarized flux spectrum, which has a 
relatively smooth, featureless continuum with spectral index $\alpha = 0$
($F_\nu \propto \nu^\alpha$). 
The polarized broad Mg II line has a FWHM of about 12,000 \kms, 
consistent with that in the total flux spectrum. 
The average Mg II FWHM in radio-loud quasars is
4620 $\pm$ 310 \kms~(with a range of 2000--10,000 \kms, Brotherton \etal~1994),
so the measured width in 3C 265 is significantly above average. 
This is most simply interpreted as 
being due to additional thermal broadening by the scattering 
plasma, a possibility that will be discussed further in \S 4.3. 
The polarization PA is featureless and constant with wavelength with a mean
of 30\arcdeg.

In addition to the strong emission lines that are seen, absorption features 
due to Ca II K at 3934 \AA~and the G band at 4301 \AA~are clearly present in 
the nuclear spectrum of 3C 265 (see Fig. 1), indicating a significant 
presence of galaxian starlight. If these absorption features are due
to a normal population of cool stars as found in typical nearby elliptical 
galaxies, then the contribution of this galaxy component to the total observed
flux is about $f_G =$ 50\% at $\sim$ 4100 \AA, dropping to to near zero at
2200 \AA, based on the galaxy subtraction method described by Tran (1995a) 
using NGC 821 as a template.
This starlight fraction is in excellent agreement with the modeling 
results of di Serego Alighieri \etal~(1996). 
The rest-frame EW of the Ca II K absorption line in the nuclear 
spectrum of 3C 265 is 4.1 $\pm$ 0.5 \AA, comparable to the value of 
5.5 $\pm$ 0.5 \AA~measured by Dey \& Spinrad (1996). 
Assuming the starlight is entirely unpolarized, we corrected the observed 
\p~for dilution, and obtained an intrinsic polarization that 
is independent of wavelength to within the uncertainties (Fig. 3c).

It is of great interest to see if similar absorption-line features are 
present in the off-nuclear spectra of the extended emission regions or
``knots." Detection of these features in the extensions would 
provide conclusive evidence for the presence of starlight being 
directly associated with these extensions, as would be expected in the
jet-induced star formation hypothesis. 
On the other hand, failure to detect these absorption features in the
optical/UV would not 
necessarily rule out this scenario since the star formation regions could
be dominated by light from hot, young stars which lack these absorption 
features. The spectra of the NW and SE extensions of 3C 265 are shown in 
Figure 4. The NW spectrum was extracted from a spatial region 
5\arcsecpoint8 wide along the slit, and that of the SE knot covered a region 
3\arcsecpoint2 wide along the slit.
These spectra do not show evidence for any absorption lines, but 
they have significantly lower S/N than the nuclear spectrum.
Higher S/N data are needed to 
conclusively rule out the presence of absorption lines.  
However, broad Mg II is detected in these spectra, consistent with 
the finding of Dey \& Spinrad (1996), and indicating that scattered
nuclear light from an obscured broad-line AGN is present in the extensions.
The high polarizations observed for these knots (Cohen \etal~1996; also see
below) strengthen this view.

Close inspection of the off-nuclear spectra of 3C 265 reveals that
the NW and the SE knots have different velocity 
profiles. In the NW the narrow emission lines are fairly symmetrical and 
redshifted relative to the nucleus, while in the SE they are 
blueshifted and have a blue-wing asymmetry, {\it plus} a weak red wing. 
Figure 5 illustrates
these differences for the profile of the [O II] \wave 3727 emission line. 
The projected velocity shifts, relative to the 
nucleus are $\sim$ +120 \kms~for the NW and $-$190 \kms~for the SE knot. 
In addition, the Balmer continuum discontinuity and high-$n$ emission lines
appear stronger in the NW spectrum than in the SE spectrum (Fig. 4). 
Detailed analysis of the line strengths is given in a separate paper, 
but we note here that these velocity shifts 
imply that if there is an outflow of material in the two sides of the 
cone, the NW lobe would be behind the SE lobe. 
VLBI observations of the inner core to look for any jet/counterjet 
flux enhancement resulting from Doppler boosting
would be useful for confirmation.
Alternatively, the line shifts could simply be due to rotation or inflow.
The origin of the line asymmetry and its difference between the two
sides are not understood, but could possibly be due to dust obscuration.

The imaging polarimetry of 3C 265 obtained through a $V$ filter 
is displayed in Figure 6. This 
shows clearly the double lobes of highly polarized radiation, 
spatially coincident with the extended UV/optical emission.
The polarization electric vectors are closely perpendicular to the 
radio axis in the inner regions close to the nucleus (for radio maps, see 
Leahy \etal~1989 and Fernini \etal~1993).
In the outer extents, although the extended emission-line 
regions (EELR) in this galaxy do not align well with the radio axis, 
the polarization vectors are well aligned with the EELR and perpendicular to 
the radius vector from the central flux peak. 
To quantify this perpendicular relationship, we examine the angle 
between a line joining the flux peak to each electric vector and the 
perpendicular to it. The mean deviation of this angle for 53 vectors 
(excluding those in a 7$\times$7 box around the nucleus which are 
contaminated by seeing and spatial averaging) is 
$-$1.4\arcdeg~$\pm$ 1.1\arcdeg. 
This remarkably tight distribution demonstrates that scattering must be
entirely responsible for the polarization, and that the nuclear source for the 
scattered light must lie close to the flux peak.
In addition, there is a general {\it increase} in $P$ with radius throughout
the spatially extended regions 
(see Table 1 of Cohen \etal~1996; see also Cimatti \etal~1996, Dey \etal~1996).
We discuss the implications of these results in \S 4.2.
Our observations confirm the results of Jannuzi \& Elston (1991)
who detected a similar level of polarization and orientation of the electric
vectors with respect to the extended optical/radio morphology.

\subsection{3C 277.2}

The spectropolarimetry of 3C 277.2 ($z$ = 0.763) is shown in Figure 7. 
Like 3C 265, the radiation is highly polarized, 
reaching $\sim$ 30\% in the blue (rest-frame UV). 
The observed polarization displays a smooth rise 
to the blue as in 3C 265, but like the latter, this is most likely due 
to the decreasing effect of starlight dilution. 
Stellar absorption lines from the host galaxy are also seen. Unfortunately, 
the Ca II K line and G band fall fortuitously on the B and A atmospheric bands,
respectively, making the determination of the relative contribution of 
the galaxy component to the total observed flux more problematic
and far less reliable. It appears, however, that the starlight fraction is 
less than that in 3C 265. 
Averaged over 4100--5100 \AA~(observed \wave), the observed polarization 
is 29\% $\pm$ 6\%, consistent with the value of 21\% $\pm$ 4\% reported by 
di Serego Alighieri \etal~(1989) in the $B$ band. 
The polarization PA is independent of wavelength, with a mean of 168\arcdeg. 
This is 73\arcdeg~from the radio structure axis of 
61\arcdeg~(McCarthy \etal~1987; see Pedelty \etal~1989 and Leahy \etal~1989
for radio maps). 
The narrow lines are essentially unpolarized as evidenced by 
the sharp drop of $P$ in these lines and their absence in \pf. 
These characteristics, including the high magnitude of polarization and 
constancy of $P$ and PA with wavelength, suggest that scattering is the 
main mechanism responsible for the observed polarization. 
\pf~appears featureless with only a smooth continuum rising slightly toward
the UV, showing no obvious sign of broad Mg II. The S/N, however is 
poorer than in 3C 265, and any broad lines present might be lost 
in the noise and difficult to detect. However, a broad component of 
Mg II is easily discernible in the total flux spectrum (Fig. 8). Fitting a
gaussian profile to this component gives a FWHM of 13,000 $\pm$ 1500 \kms, 
similar to that of 3C 265, and indicating that it too could be thermally 
broadened by the scatterers. 

Like 3C 265, the spectrum of 3C 277.2 shows emission lines of a wide 
range of ionization and atomic species. Most of the lines
identified in 3C 265 are observed in 3C 277.2, and
their measured line ratios are listed in Table 2. 
The weakness of any broad line components
in \pf~prevents us from estimating the strength of any unpolarized
featureless continuum component FC2, 
but it appears (due to the higher \p) that any FC2 present is weak 
compared to that in 3C 265 (see \S 4.1). These results suggest that 3C 277.2
represents another case of a quasar hidden from direct view
in a normal radio galaxy, and thus support the radio galaxy/quasar
unification hypothesis (Barthel 1989).

A polarization map of 3C 277.2 obtained with a $V$ filter is shown in Figure 9. 
The EELR is less extended than in 3C 265, but the polarization
is high near the nucleus. 

\subsection{3C 324}

Keck spectropolarimetry of 3C 324 ($z = 1.206$) has recently been discussed
by Cimatti \etal~(1996). 
We present the $B$-band imaging polarimetry of this object in Figure 10. 
The polarization level is high throughout much of the nuclear and 
extended emission regions, with PA roughly perpendicular to the 
radio/extended optical structure axis (VLA radio maps of the source can be
found in Pedelty \etal~1989 and Fernini \etal~1993). 
The level of the observed polarization in the nucleus 
(12\% $\pm$ 1.3\%) is consistent with the measurements of 
Cimatti \etal~(1996, $\sim$ 11\%). 
Again, the rise in \p~in the outer parts of the galaxy is present, 
especially towards the east, similar to that observed in 3C 265. 
These results 
confirm those reported by Cimatti \etal~(1996) with their long-slit 
data. The apparent ``extended lobe" $\sim$ 3\arcsecpoint3 to the west 
of the nucleus is essentially unpolarized. 
%although it is of about the same brightness as the nucleus of 3C 324. 
It is most likely an interacting/merging or companion galaxy at a redshift
similar to that of 3C 324 (Cimatti \etal~1996).

\subsection{3C 343.1}

In contrast to 3C 265 and 3C 277.2, the nuclear flux spectrum 
of 3C 343.1 ($z = $ 0.750, Fig. 11)
is characterized by the weakness or absence of higher ionization 
lines like [Ne V] or He II and by the presence of easily
detectable broad wings at H$\beta$, H$\delta$, and probably Mg II. 
In addition, the 
absorption line spectrum is dominated by a younger population of stars,
as can be seen by the prevalence of absorption lines due to high-$n$ Balmer
lines, the Balmer discontinuity, and the lack of a feature due 
to the G band that is characteristic of cooler stars seen in 
the 3C 265 and 3C 277.2 spectra. 
A most surprising finding, however, is that the redshift of the absorption
lines is radically smaller ($z=0.344$) than that of the nuclear 
AGN emission lines ($z = 0.75$) ! 
As seen in Figure 11, the low-redshift system also contains emission lines, 
notably [O II] \wave 3727 and \oiii~\waves 4959, 5007. 

Examination of the 2-dimensional spectroscopic data reveals that 
the [O II] \wave 3727
emission line in the low-$z$ emission-line galaxy is extended over 
$\sim$ 5\arcsec, while the emission lines of [O II] and Mg II from the
active nucleus are relatively compact. This demonstrates that the observed 
spectrum is composed of two different sources of
radiation in a chance alignment. 
The distant object with $z = 0.75$ is the radio AGN 3C 343.1. 
The foreground object with $z=0.344$ has a spectrum 
similar to that of the emission-line galaxy M51, showing 
the strong emission line
of [O II], as well as strong absorption lines due to Ca II H \& K and 
high-$n$ Balmer lines (Fig. 11). 

Based on the strength of [O II] relative to [O III], the foreground galaxy
could be classified either as a LINER (low-ionization nuclear emitting
region) or starburst galaxy, although a measurement of [O I] \wave 6300, 
which is lost in the high-noise region of the spectrum in the near-IR,
is needed to establish its true classification. 
Nevertheless, the spectrum qualitatively appears much like that of 
M51, and we use the latter as a template to estimate 
the contamination to the nuclear spectrum of 3C 343.1 by the
foreground galaxy. 
We find that the foreground galaxy contributes 
about 50\% of the flux at the wavelength of Mg II, rising to nearly 70\% 
at the wavelength of \hb. Figure 12 shows the spectral decomposition. 
Thus, contamination of the 3C 343.1 spectrum is substantial, and any study
of its intrinsic optical continuum must take this into account before 
interpretation can be made.  

3C 343.1 is classified as a compact steep spectrum 
radio source (CSS) (see, e.g., Fanti \& Spencer 1996). 
The VLA and VLBI radio images show a compact double structure with the
two components separated by
$\sim$ 0.3--0.5\arcsec, extended east-west (van Breugel \etal~1992; Fanti 
\etal~1985). No polarization is detected at either 8.4 or 15 GHz 
(van Breugel \etal~1992; Akujor \& Garrington 1995). 
In light of the fact that 3C 343.1 is closely aligned with a foreground
galaxy at $z=0.344$, the double morphology of the radio source 
could be a result of gravitational lensing. We note, however, 
that such morphology is rather common for CSS objects
(see, e.g., Fanti \etal~1985), and no lensing may be involved. 
However, if this is a lens system, given the redshifts
and radio separation, the mass of the foreground galaxy is estimated
to be $\sim$ 3.5 $\times 10^9$ M$_{\sun}$~(Fassnacht 1996, private communication).
The similarity of the radio intensity ratios at 15 GHz and 22.5 GHz between 
the two components in the VLA map (van Breugel \etal~1992) is consistent 
with their being a lensed system.
Matching VLBI-scale structure and spectra in the radio images
should be important in establishing whether or not this is indeed the case.
The VLBI image of Fanti \etal~(1985), however, shows very different shapes
between the two components, so there is no
strong evidence for lensing in the 3C 343.1 system. 
A recent {\it Hubble Space Telescope} ($HST$) snapshot survey of CSS radio
sources by de Vries \etal~(1997) does not reveal any peculiar structures in 
the field of 3C 343.1. Deeper $HST$ imaging would be most 
informative and should help resolve the true structural morphology of 
this system. 

Imaging observations in the light of the optical continuum and 
[O II] emission of the AGN by McCarthy \etal~(1995) show a continuum 
structure slightly extended but offset 45\arcdeg~from that 
of the [O II] morphology, which is elongated east-west in the direction of
the radio axis. The AGN [O II] emission has a $\sim$ 5\arcsec~extent, 
similar to the extension of the [O II] emission in the foreground galaxy. 
Since we have shown that the optical continuum of 3C 343.1 is dominated by 
a foreground emission-line galaxy of similar spatial extent, it is easy 
to understand the observed misalignment between the continuum 
and [O II] morphologies of the radio galaxy, and also 
why the optical emission in 3C 343.1 is extended well beyond the region
of the radio emission (McCarthy \etal~1995). This property is not unique 
to 3C 343.1, but it is an extreme example. 
Because of the large distances and faint limiting magnitudes of these HzRGs, 
the chance alignment with a foreground source does not appear to be an 
uncommon occurrence. For example, 
the projection of a Galactic M dwarf within 1\arcsec~of the nucleus 
of 3C 368 has been reported by Hammer, Le F\`evre, \& Proust (1991), and
the superposition of a Galactic star with 3C 435B
has been discussed by McCarthy, van Breugel, \& Spinrad (1989).
Furthermore, as there has been statistical evidence for an excess of 
foreground galaxies around HzRGs, presumably due to amplification bias 
caused by gravitational lensing (e.g., Hammer \& Le F\`evre 1990; Ben\'itez,
Mart\'inez-Gonz\'alez, \& Martin-Mirones 1997), chance superpositions
of these objects are to be expected. 

We detect little polarization in the nucleus of 3C 343.1. The observed
$P$ integrated over the wavelength range 5000--7600 \AA~(observed) is 
0.6\% $\pm$ 1.6\% (uncorrected for the $z = 0.344$ galaxy). 
This is consistent with the upper limit of $<$ 9\% reported by 
Tadhunter \etal~(1992), but it is also consistent with zero.
The presence of broad \hb, and lack of [Ne V] and He II, 
coupled with the observation that it is weakly or not polarized at 
all would suggest that 3C 343.1 is a low-ionization broad-line 
radio galaxy viewed at a small inclination angle. 
The heavy contamination with a foreground, young emission-line galaxy further 
contributes to its lack of observed polarization. 
3C 343.1 is also known as a CSS radio source, the polarization properties 
of which as a class have not been well studied, but recent evidence 
(Cohen \etal~1997) points to the intriguing possibility that the optical
polarization of these sources comes from synchrotron radiation and is variable.

\section{Discussion}
\subsection{Total Flux Spectra and FC2}

In 3C 265 the EW of the broad Mg II line is found to be 
significantly higher in \pf~than in the total flux 
[58 $\pm$ 3 \AA~compared to 12 $\pm$ 1 \AA~(rest frame), 
respectively]. As mentioned
by Cohen \etal~(1996), this can be interpreted as being due to the 
presence of an unpolarized featureless continuum component, FC2, in the 
observed spectrum.  
This Mg II EW in \pf~is similar to those observed directly in normal 
quasars (Cristiani \& Vio 1990), whereas in the hybrid Doppler-beamed 
continuum model, one would expect it to be much lower.
Thus, any reflected Doppler boosted component must be minimal in 3C 265.
We can estimate the fraction of FC2 as (e.g., Tran 1995c):
FC2/$F_{tot}$ = 1 $-$ EW($F_{tot}$)/EW(\pf) = 0.79. The intrinsic
polarization of the scattered component is then about 50\%. 
Since the starlight component from an old stellar population of the 
host galaxy is expected to be insignificant at this wavelength, 
this FC2 appears to be related 
to, or the same as that recently found in an increasing number of Seyfert 2 and 
radio galaxies (Tran 1995c; Tran \etal~1995; Dickson \etal~1995; 
Dey \etal~1996; Cimatti \etal~1996).
An interesting possibility is that FC2 is due to hot
stars being formed in the wake of the radio jets in the extranuclear regions
(Longair, Best, \& R\"ottgering 1995; Best \etal~1996). 
A radio jet interacting with the circumnuclear medium has been thought to 
induce star formation (e.g., Rees 1989), and this has been proposed to 
explain the alignment effect. 
If this is the case, hot stars in the extended regions could contribute 
a significant  amount of UV light which could account for 
much, if not all, of the FC2. 
Signatures characteristic of stars in this 
spectral regions, such as the spectral breaks near 2600 \AA~and 
2900 \AA~(Fanelli \etal~1992) could provide crucial information 
on this issue. However, such signatures are fairly weak for O, B type 
stars, and the additional presence of AGN light makes them difficult to detect
conclusively. Searches for these features have been made in the spectra 
of 3C 256 and 3C 324, with inconclusive results. 
The observed continua are consistent both with a young stellar population, 
and also with a nonstellar continuum (Dey \etal~1996; Cimatti \etal~1996).
Similarly, the spectra of the off-nuclear knots in 3C 265 
(Fig. 4) also show no significant sign of these breaks, consistent with
either hot stars or nonstellar light.

Moderately hot stars (A type) could make their presence
known from the high-$n$ Balmer absorption lines. No such features 
are evident in the spectrum of 3C 265, although they may be impossible to 
detect because of the overlying Balmer emission lines.
Young A-type starlight has been observed in 3C 343.1, but as shown in \S 3.4, 
it comes from a foreground galaxy, not the host.
However, many absorption features commonly associated 
with early type galaxies, such as Ca II H \& K, 4000 \AA~break, 
and the G band are clearly seen (see Fig. 1).
These features demonstrate the presence of cool, evolved stars, 
and thus the host galaxies of these sources have a ``normal" character. 
Recent $HST$ imaging of nearby quasars by Bahcall, Kirhakos, 
\& Schneider (1995, 1996) has shown that 
luminous quasars occur in a great variety of host systems: from 
apparently normal elliptical and spiral galaxies, 
to complex interacting components. 
If the hosts of these HzRGs are similar to those of quasars in the Bahcall
sample, as may be the case since many radio galaxies have been shown 
to be quasars viewed from an unobstructed direction (e.g., this paper; 
Tran \etal~1995; Young \etal~1996; Dey \etal~1996; Cimatti \etal~1996, 1997),
then the finding that some of them appear to be normal, evolved systems is 
consistent with the Bahcall \etal~observations.

Alternatively, FC2 could simply represent the nebular continuum generated 
in situ in the emitting gas (Dickson \etal~1995) or it could be 
free-free thermal 
radiation from the scattering electrons themselves (Tran 1995c). 
For the latter to be valid, the scattering medium must be dominated 
by electrons and not dust. Nebular continua are detected in both
the nuclear and off-nuclear (NW) spectra of 3C 265 (Figs. 1, 4) and in
the spectrum of 3C 277.2 (Fig. 7), based on the presence of the Balmer 
discontinuity around 3650 \AA. 
Careful examination of the scattering mechanism 
should provide important insights into the origin of this FC2 component.
The recognition of FC2 in these galaxies means that it could make up part
of the observable off-nuclear emission regions.
If so, determination of the origin of this light would be of great value in 
understanding the alignment effect and the evolution of these galaxies, as well
as the nature of quasars in general. 
For example, $HST$ images of the nearby 
luminous quasar PKS 2349-014 by Bahcall \etal~(1995) have
revealed a region of very diffuse and extended ``nebulosity" surrounding the
nucleus, and extending up to $\sim$ 40 pc. 
The origin and nature of this light is unknown. 
Similarly, an $HST$ snapshot survey of Seyfert galaxies 
has revealed extended ``fuzz" surrounding the nuclei of 
type 2, but not type 1 Seyfert galaxies (Nelson \etal~1996).
If the quasars at the centers of these HzRGs are similar to 
PKS 2349-014, this faint extended nebulosity could represent a component of
FC2 that dilutes the polarization. It could, however, be tidal debris from
galaxy interactions (e.g., Weil, Bland-Hawthorn, \& Malin 1997). 
Further study of the host galaxies and FC2 of these HzRGs will
shed light on the nature of this nebulosity in quasars. 

\subsection{Scattering vs. Star Formation}

In order to further investigate the geometry of the surrounding environments of 
HzRGs, and to help address the question of scattering vs. star formation, 
imaging polarization observations were obtained for 3C 265, 3C 324 and 
3C 277.2. The results, shown in Figures 6, 9 and 10, reveal a number of 
interesting characteristics. First, the polarizations are all very high,
and consistent with the nuclear spectropolarimetry. 
Second, the polarization vectors, especially in 3C 265 (\S 3.1), show a 
double-lobed pattern that is centrosymmetric with respect to the nucleus, and 
perpendicular to the radial vectors originating from it. 
Such bisymmetric fans of polarized light have also been observed in the
nearby powerful NLRG Cygnus A (Tadhunter, Scarrott, \& Rolph 1990; 
Ogle \etal~1997).
This is just what is expected from a simple geometry in which radiation 
from a compact obscured central source is scattered by the surrounding medium. 

It is notable that the polarized light is spatially coincident with, 
and the polarization PA more closely orthogonal to the UV extensions than 
the radio axis, as also discussed by Cimatti \etal~(1996, and references 
therein) for many other radio galaxies. 
This phenomenon in fact, has also been seen in Seyfert galaxies. 
For example, Tran (1995b) found that the polarization vectors in the 
luminous hidden quasar Mrk 463E are more perpendicular to the optical-UV 
extension (Uomoto \etal 1993) than to the radio structure axis, suggesting 
that the axis of symmetry of the 
scattering material lies along the optical extension and not the radio jet.
The similar phenomenon in HzRGs suggests that scattering of light
by material closely associated with the UV extensions rather than 
the jet is responsible for the observed perpendicular relationship. 
Moreover, it is interesting to note
that in the case where the radio axis and UV extensions do not align well 
(e.g., 3C 265), this relationship also applies more 
with the UV extensions (see Fig. 6).
This result clearly favors the scattering picture over jet-induced 
star formation as the main cause of the alignment effect. 
Di Serego Alighieri \etal~(1996) have argued that the 
apparent misalignment between the radio axis and UV extensions in 
3C 265 considerably weakens the case for 
star formation hypothesis, since in this picture, it is hypothesized that
the radio jets trigger the star forming process and one would therefore
expect a very close correspondence between the optical and radio structures 
in all cases. 

The third notable characteristic displayed by the polarization images 
is that \p~shows a spatial gradient (most clearly seen in 3C 265) 
in the sense that its magnitude increases with increasing radial 
distance from the nucleus. 
This rise in the degree of polarization in the outer regions of HzRGs
away from the nucleus has also been noted in both 3C 256 and 3C 324 
by Dey \etal~(1996) and Cimatti \etal~(1996) in their 
one-dimensional spectropolarimetric data. 
Thus, it appears that this phenomenon is wide-spread 
if not universal among these objects. 
If the intrinsic polarization of the scattered light at every point in 
the galaxy is the same, this gradient in $P$ may simply be due to 
the decreasing effect of dilution by diminishing unpolarized starlight 
at the outer regions of the galaxy, a characteristic suggested
by Draper, Scarrott, \& Tadhunter (1993). 
If radiation from luminous, young stars in their forming 
regions dominated in the extended emission regions, the 
opposite effect would be expected.
Furthermore, it is unlikely that the gradient is caused by diminishing 
light from cool stars since the effect is seen in the observed $V$ and $B$ 
bands ($\sim$ 3000\AA rest-frame), where radiation from such stars is not 
expected to dominate (e.g., Fig. 1, 3). The ramification is that scattered 
light, not direct starlight must increasingly dominate with radius in the 
extended regions. Again, this provides strong support for the scattered light 
hypothesis in explaining the alignment effect. 

Another possibility is that near the nucleus, scattering takes place in
an extended (3-dimensional) volume, so there is some cancellation of the 
\p~vectors and the integrated polarization is reduced. At large distances, 
the scattering occurs mainly through large angles, and the polarization 
is correspondingly higher.
Alternatively, the radial increase in $P$ 
could be the result of a radial increase 
in scattering efficiencies and/or number of scatterers. 
How such differences in the scatterers' 
properties could arise is not known. 
One intriguing possibility is that close to the nucleus the polarization
arises mainly from electron scattering, but farther out, it is dominated
by dust scattering, which has higher scattering efficiency. 
An attractive feature of this picture 
is that it has been shown to operate in the nearby 
Seyfert galaxy NGC 1068 (Miller \etal~1991). 
In this case, however, diagnostic signatures for dust scattering are 
clearly seen, while they have not been convincingly demonstrated in 
HzRGs (see, however, Knopp \& Chambers 1997). 
In summary, imaging polarimetry provides convincing evidence that
scattered radiation from an obscured nuclear source rather than 
star forming regions dominates the extended emission in the HzRGs of this
study.

\subsection{Nature of the Scatterers}

To address the question of the nature of the scatterers, one can first attempt 
to look at the intrinsic wavelength dependence of \p.  
In \S 3.1 we showed that the intrinsic polarization of 3C 265 is essentially 
wavelength independent, and that the apparent rise of $P$ toward the blue 
is due entirely to decreasing dilution by unpolarized starlight (Fig. 3c).
The polarization PA is also wavelength independent. 
It is important to note that Dey \etal~(1996) and Cimatti \etal~(1996, 1997) 
also find that $P$ and PA are essentially independent of wavelength 
in 3C 13, 3C 256 and 3C 324. 
%Normal Galactic-type dust grains with an MRN distribution would produce
%\p~that rises sharply to the blue because of the rapidly 
%increasing scattering cross section with increasing frequency (White 1979).
Although this is consistent with electron scattering, scattering by dust 
grains with the appropriate characteristics can just as well mimic the 
constancy of \p~and PA with wavelength (e.g., Kartje 1995; 
Manzini \& di Serego Alighieri 1996). 
It is not possible, therefore, in the limited optical wavelength range, to 
discriminate between electron and dust scattering by examining the 
wavelength dependence of \p~alone.

One might expect that dust scattering may play 
a significant role in producing the observed polarization in these HzRGs.
There has been evidence that dust must surely be present
in many of these radio galaxies (Heckman, Chambers, \& Postman 1992; Chini \&
Kr\"ugel 1994; Dey, Spinrad, \& Dickinson 1995; Ivison 1995; Villar-Mart\'in \& Binette 1996, 1997), 
although there is no direct information on how it is spatially distributed.
In addition, dust is much more efficient at scattering the light than 
electrons (by a factor of order $\sim 10^4$, in terms of scattering cross 
section per unit mass). 
In fact, in many cases it has been thought to be the main or dominant 
process responsible for the observed polarization (Cimatti \etal~1993; 
di Serego Alighieri \etal~1994; Manzini \& di Serego Alighieri 1996; 
Knopp \& Chambers 1997).

Some arguments against dust include the lack of any sharp rise in $P$ in 
the UV for $\lambda \simlt$ 2500 \AA, as would be expected 
from theoretical predictions
regardless of the grain size, composition, and distribution 
(Kartje 1995; Manzini \& di Serego Alighieri 1996).
In addition, observable signatures that are indicative of dust scattering, 
such as the extreme blueness 
($\alpha \sim +2$) of the polarized, scattered continuum 
[Cygnus A (Ogle \etal~1997), NGC 7674 (Miller \& Goodrich 1990; 
Tran 1995b)] and the lack of any significant broadening of permitted
lines in \pf~spectrum [NGC 7674 (Miller \& Goodrich 1990; Tran 1995b), 
the NE knot in NGC 1068 (Miller \etal~1991)],
have not been seen in many HzRGs observed so far.
The spectral index of \pf~in 3C 265 is $\alpha = 0$ and in 3C 277.2 
$\alpha = -0.2$. These values are consistent with the typical spectral 
slopes of normal QSOs (Francis \etal~1991; Natali \etal~1998) with no 
significant ``bluening," suggesting that some type of grey scatterers 
may more likely be responsible. The lack of any observable polarization or 
spectral flux signatures associated 
with the 2175 \AA~``bump" of the interstellar extinction curve is also 
puzzling, although the composition, size distribution and extinction 
properties of the dust in these galaxies could be very different from 
those in our Galaxy.
For example, the extinction curves of the Small Magellanic Cloud 
and of starburst galaxies lack such a 2175 \AA~dust feature 
(Pr\'evot \etal~1984; Calzetti, Kinney, \& Storchi-Bergmann 1994; 
Gordon, Calzetti, \& Witt 1997; Rodrigues \etal~1997). 
Also, Manzini \& di Serego Alighieri (1996)
have argued that it is possible to reduce the strength of this signature in 
the scattered flux by introducing an appropriate amount of extinction in 
addition to scattering. Such a process may also help explain the lack of 
any strong bluening in the observed scattered spectra. 

Electrons are thought to be the dominant scattering agent
in Seyfert galaxies\footnote{The situation for Seyfert 1 galaxies is 
less clear, but both electrons and dust are thought to contribute (Goodrich
\& Miller 1994).} (e.g., Miller \etal~1991; 
Tran 1995b), albeit at spatial scales much closer to the nucleus 
($\simlt$ a few pc) than those associated with the HzRGs described here 
($\sim$ tens of kpc). 
One clue that argues in favor of electron scattering in 3C 265 
and 3C 277.2 is the broadening of the Mg II line in \pf~(\S 3.1, 3.2). 
The magnitude of the broadening is such that if 
electrons were the scatterers, the temperature of the plasma clouds is
$\sim$ 5 $\times$ 10$^5$ K, not hot enough to smear out
the line completely but comparable to the temperature found
for the scattering electrons in NGC 1068 (Miller, Goodrich, \& Mathews 1991). 
Of the previous HzRGs that have been observed, we note
that in 3C 256, 3C 324 and 3C 356 electron scattering could not be ruled 
out by the data and might even be slightly favored in 3C 256 
(Jannuzi \etal~1995; Dey \etal~1996; Cimatti \etal~1996, 1997).

This leads us to conclude that while there is no doubt that dust is present 
in these galaxies and most probably responsible to some extent 
for the scattered light, electron scattering also contributes significantly,
despite its lower scattering efficiency. 
The only main argument against electron scattering is the uncomfortably large
mass of ionized gas ($\sim 10^9 - 10^{12}$ M$_{\sun}$, 
see, e.g., Dey \etal~1996) that would be required, due to the large 
spatial scales involved.
Nevertheless, if electron scattering is the dominant process, it 
is difficult to attribute the alignment effect to 
star forming regions alone, as one would expect scattering by dust that is 
abundant in the molecular clouds from which stars form might dominate.
As discussed by Dey \etal~(1996), another consequence of the electron-dominated 
scattering picture is the implication for a dust-to-gas ratio
in the interstellar medium of these galaxies that is smaller than that 
in our Galaxy. Although the issue is far from resolved, recent evidence 
suggests that the amount of dust in high-redshift galaxies 
(Pettini \etal~1997) appears to always be 
smaller than in the Milky Way, consistent with the above conclusion.

\section{Summary and Conclusions}

The main observational results of this paper are: 

1. Broad permitted emission lines (Mg II) are observed in \pf~(3C 265) and 
total flux (3C 265, 3C 277.2). 

2. The widths of the Mg II lines in \flam~(3C 265 and 3C 277.2) and
\pf~(3C 265) are extraordinarily large compared to normal, radio-loud quasars.
A similar broadening is also observed in 3C 324 and 3C 356 (Cimatti
\etal~1996, 1997).

3. The $P$ and PA are generally constant with wavelength, after 
dilution by host starlight is taken into account. 
This appears to be a common phenomenon in many HzRGs (e.g., 3C 13, 3C 277.2,
3C 256, 3C 265, 3C 324).

4. Polarization PAs are perpendicular to the UV extensions, and perhaps 
more importantly, to the overall radial structure of the galaxy. 
$P$ levels are generally high, and comparable to those in Seyfert galaxies. 

5. Imaging polarimetry of 3C 265 shows that the $P$ vectors display 
a centrosymmetric pattern with respect to the nucleus, with magnitudes 
that generally increase radially outward from the nucleus into the 
extended emission regions. 

6. Spectroscopically, $P$ shows a slight increase in the wings of broad 
emission lines (e.g., Mg II) compared to the continuum or equivalently, 
the EW(\pf) $>$ EW(total flux), as has been seen in Seyfert 2 galaxies 
and the radio galaxy 3C 234 (Tran 1995c; Tran \etal~1995). 

7. Narrow emission lines are unpolarized, suggesting that the scattering 
material is primarily inside of, or coincides spatially with the emitting 
gas of the EELR. 

8. Emission-line spectra of 3C 265 and 3C 277.2 are rich in a wide range 
of ionization and atomic species. Nebular continua are detected in the
nuclear and off-nuclear spectra of 3C 265 and 3C 277.2.
No stellar absorption lines are detected in the 
off-nuclear extensions, although the present S/N is poor. Better 
quality data are needed to examine this issue more closely and 
provide a definitive statement.

9. Host galaxies in HzRGs appear to be normal (evolved) elliptical galaxies 
or spiral bulges with a population of cool, evolved stars, based on the 
success of a normal elliptical template in decomposing the observed spectra. 

10. 3C 343.1 is found to be a superposition of two galactic systems
along the line of sight; one is a radio AGN at high redshift, and the other
a foreground, emission-line galaxy containing a young stellar population. 

Results 1--9 provide strong support for the hidden quasar 
hypothesis and the unification of radio galaxies and quasars based
on orientation. 
They indicate the presence of an unpolarized extended continuum component
of the UV light, FC2, which 
has a decreasing contribution in the extensions, causing \p~to rise there
compared to the nucleus. Together with the centrosymmetric fan of 
polarized light around the nucleus, this clearly shows that scattered light
from a central obscured source dominates the extended emission regions, 
and argues against the jet-induced star formation hypothesis as the main 
cause of the alignment effect. 
Scattering of Doppler-boosted light from a relativistic jet cannot 
be the dominant effect in these objects 
since we see broad Mg II in \pf~with an EW comparable with that observed 
directly in normal (non-blazar) quasars. 
FC2 could be hot stars, the nebular continuum or another component that
might be related to the ``diffuse nebulosity" observed in $HST$ images of 
quasars. 
Another possibility to explain the radial \p~gradient is that the scattering 
in the nucleus is dominated by electrons 
(less efficient), while in the outer extensions, 
it is dominated by dust (more efficient). However, 
no common optical/UV polarimetric signatures for the
latter, such as a rapid rise of $P$ with shorter 
wavelength or an extremely blue polarized continuum, as observed 
for the types of dust grains in our Galaxy and other Seyfert galaxies, 
have been seen in HzRGs as a class.
While previous works have suggested evidence for a large amount of dust 
in HzRGs, and scattering by this dust cannot be entirely ruled out by 
polarimetric data, there is evidence that electron scattering plays a 
significant and perhaps major role in producing the polarized light observed 
in these galaxies. 
Along with the studies of Dey \etal~(1996) and Cimatti \etal~(1996, 1997), 
these conclusions apply well to those HzRGs at $z \sim$ 1 that are highly 
polarized. There are indications that at much higher redshifts the situation 
may be very different, with star forming regions dominating the extended light 
(e.g., Dey \etal~1997), implying cosmological evolution of distant 
radio galaxies.

\acknowledgments

The W. M. Keck Observatory is operated as a scientific partnership between
the California Institute of Technology and the University of California. It
was made possible by the generous financial support of the W. M. Keck
Foundation. We are grateful to Tom Bida for assistance with 
the observations, and to A. Martel for assistance with some of the
reduction. We thank D. E. Osterbrock for useful discussions and his
help with the identification of some of the emission lines.
We thank an anonymous referee whose suggestions clarified several points.
Work performed at the IGPP/LLNL is supported by the DOE under contract
W7405-ENG-48.
During the course of this work, H.D.T. was supported by postdoctoral 
research fellowships at the California Institute of Technology through 
NSF grant AST-9121889, and at Lick Observatory through grant AST-8818925. 
S.S.A. thanks Caltech for hospitality during this work.
This research has made use of the NASA/IPAC Extragalactic
Database (NED) which is operated by the Jet Propulsion Laboratory,
Caltech, under contract with the National Aeronautics and Space Administration.

\clearpage

\begin{figure}
\plotone{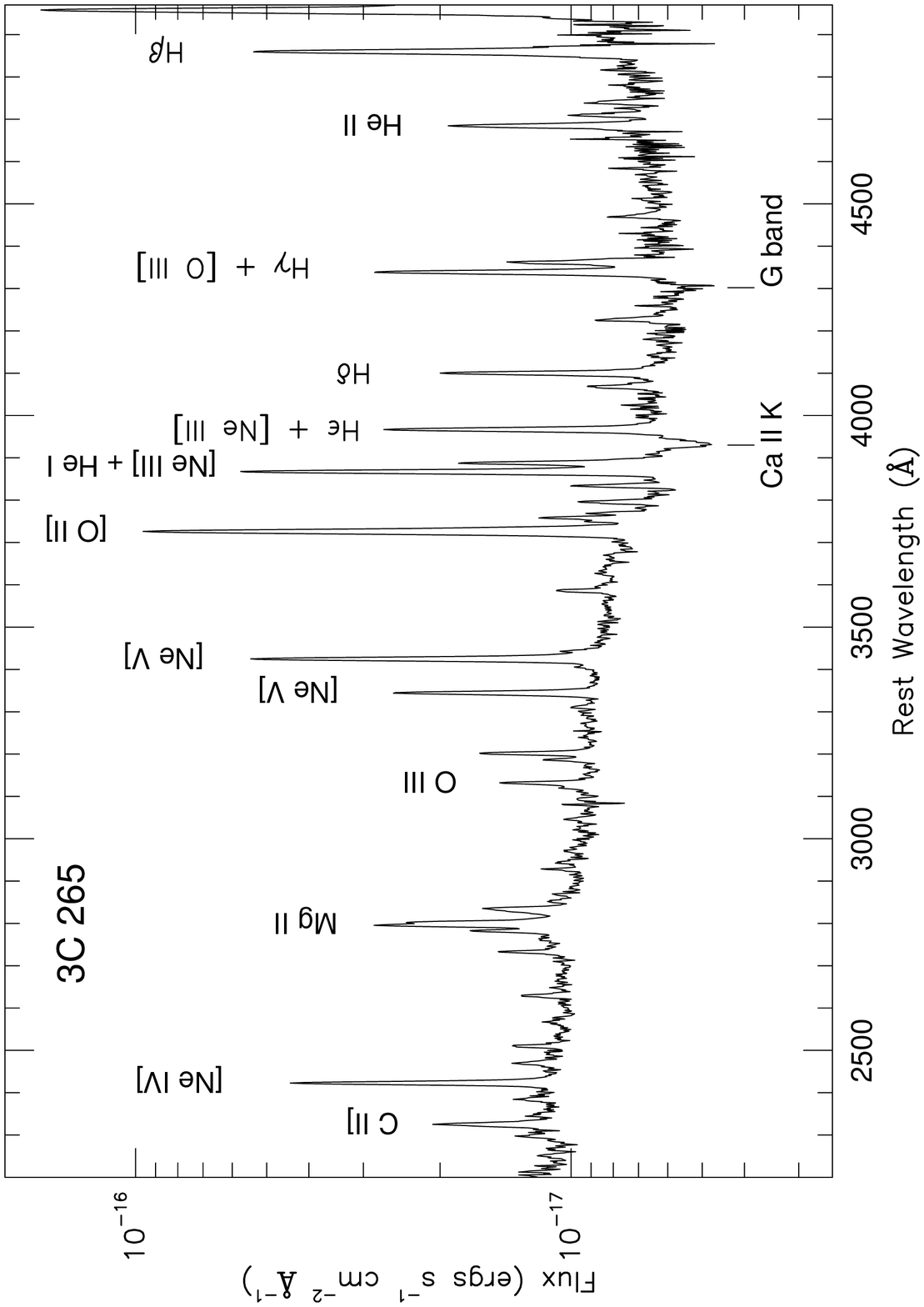}
\caption{
Total flux spectrum of the nucleus of 3C 265, plotted in
logarithmic scale to show faint features. Note the presence of 
broad Mg II emission, and the stellar absorption features. \label{fig1}}
\end{figure}

\begin{figure}
\plotone{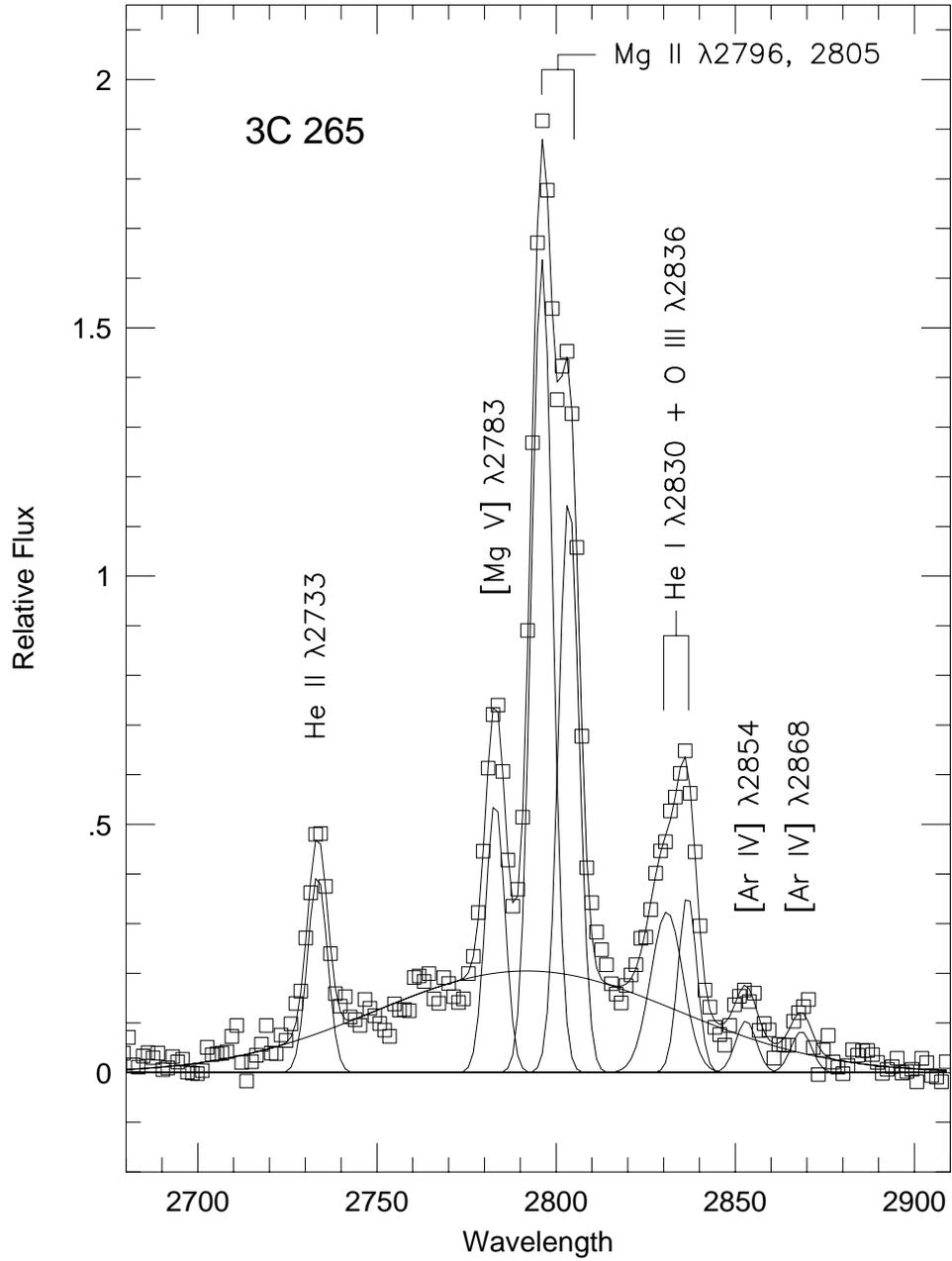}  
\caption{Fitting of Gaussian profiles to the Mg II emission-line
complex of 3C 265 in total flux. The broad Mg II line has  
FWHM = 11,000 $\pm$ 700 \kms, consistent with that in \pf. \label{fig2}}
\end{figure}

\begin{figure}
\epsscale{0.9}
\plotone{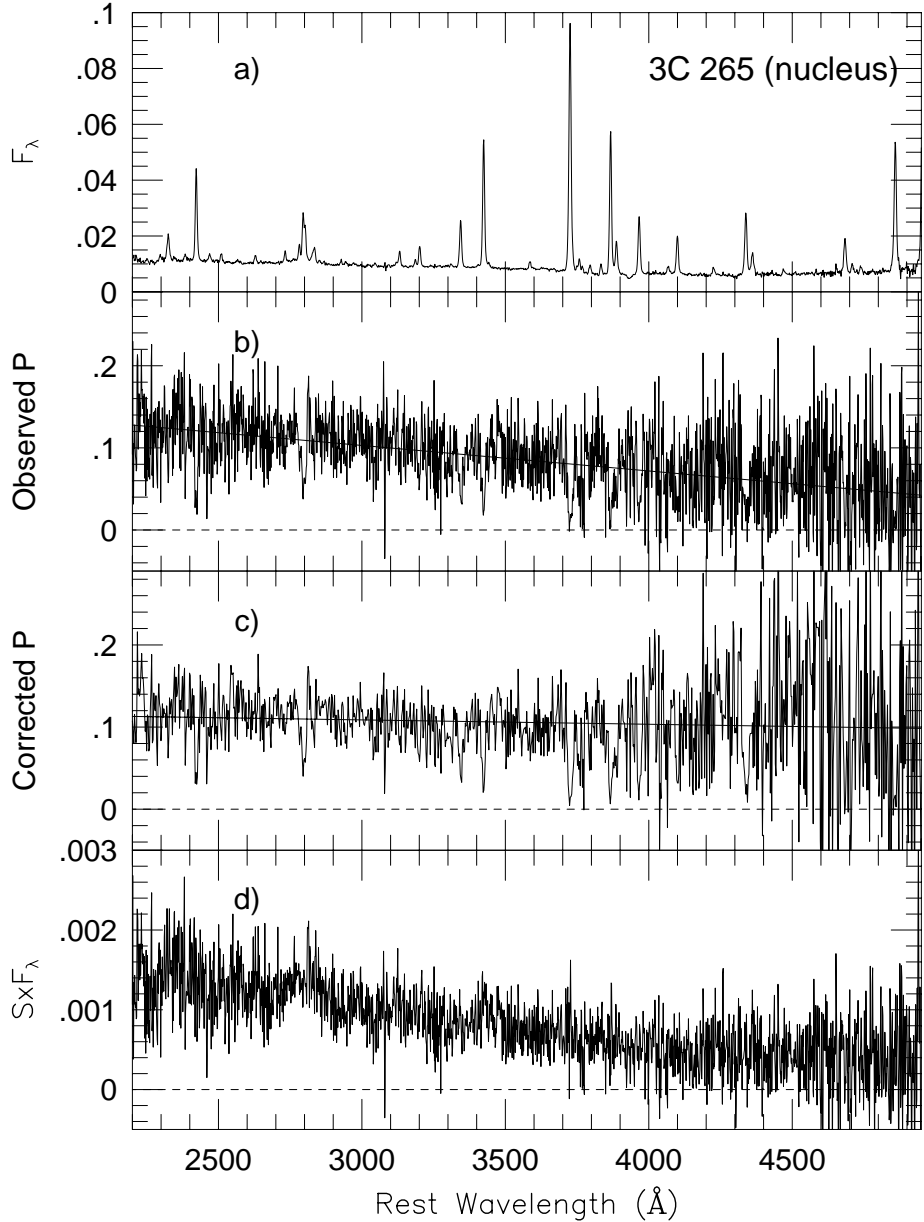}  
\caption{Spectropolarimetry of 3C 265 nucleus. 
From top to bottom are
the (a) total flux spectrum \flam, (b) observed degree of polarization \p, 
(c) \p~after correction for dilution by galactic host starlight, 
and (d) polarized flux spectrum \pf. 
The flux scales are in units of 10$^{-15}$ ergs s$^{-1}$ cm$^{-2}$ \AA$^{-1}$.
The smooth curves in (b) and (c) are straight-line fits to the continuum 
polarization. Note the presence of broad Mg II and lack of narrow emission 
lines in \pf. Note also that the increase in the observed \p~toward shorter 
wavelengths is flattened out in the corrected \p~spectrum (c). \label{fig3}}
\end{figure}

\begin{figure}
\plotone{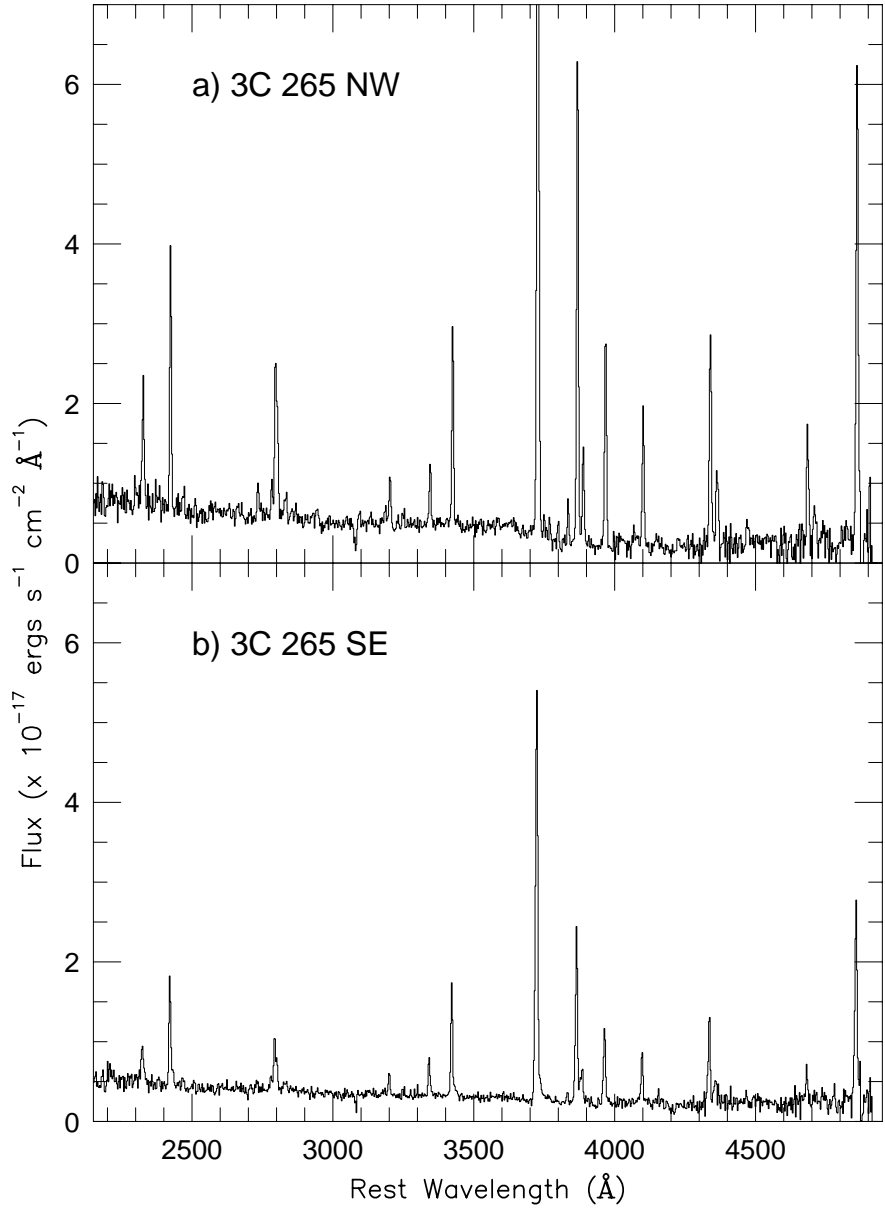}  
\caption{Off-nuclear spectra of the NW and SE emission-line extensions 
of 3C 265. (a) The NW extension; (b) the SE extension. 
Note the lack of the Ca II K and G band stellar absorption feature, 
indicating the lack of a concentration of old stars in these regions. 
Broad Mg II is present at a low level. \label{fig4}}
\end{figure}

\begin{figure}
\plotone{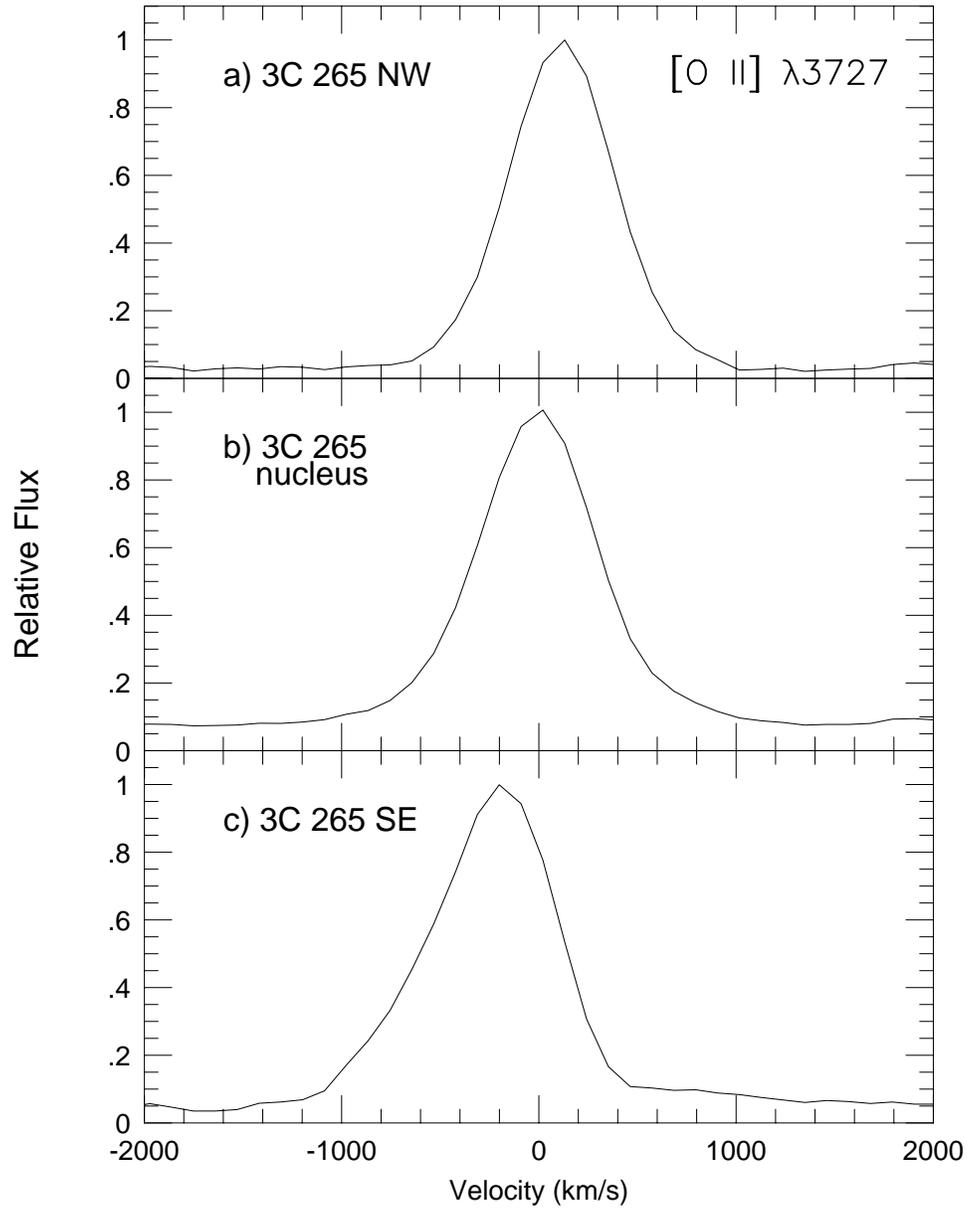}  
\caption{Emission-line profile of [O II] \wave 3727 in the (a) NW, 
(b) nucleus, and (c) SE regions of 3C 265 to illustrate the differences
in velocity profiles and shifts among the three components. \label{fig5}}
\end{figure}

\begin{figure}
\epsscale{0.85}
\plotone{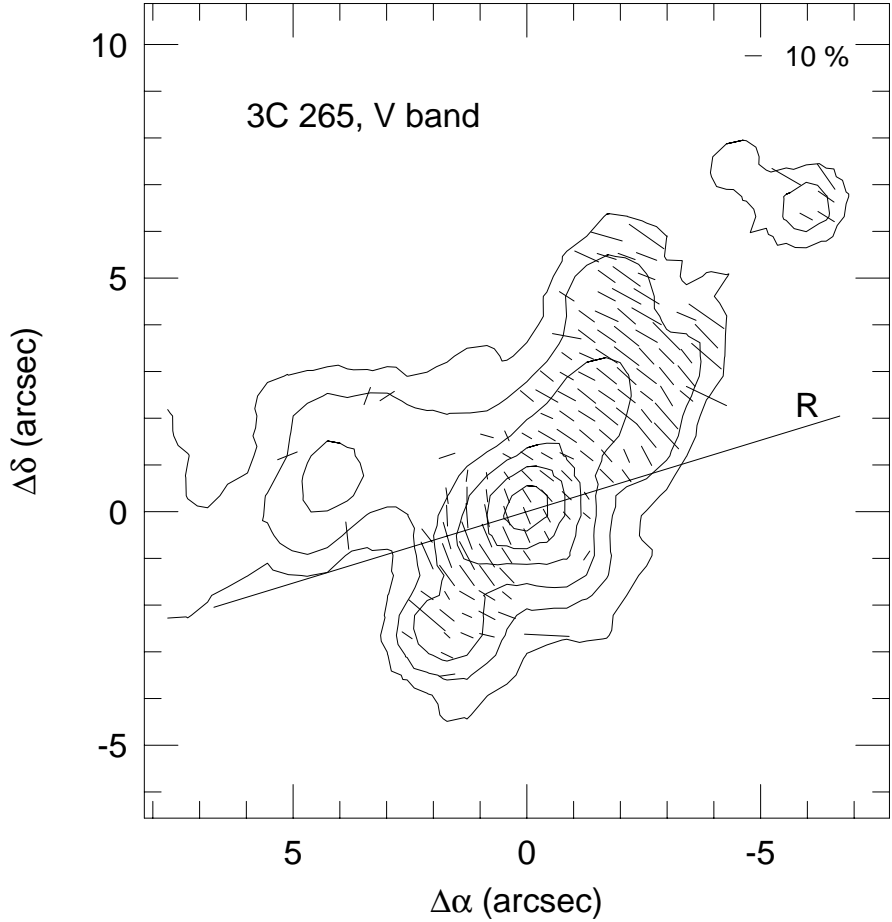}  
\caption{$V$-band imaging polarimetry of 3C 265. Polarization vectors
are superimposed on the total intensity contours. 
Contours are 2, 4, 8, 16, 32, 64\% of peak intensity.
In this and all subsequent polarimetric images, only vectors whose
S/N $>$ 2 are plotted; north is up and east is to the left. 
The data have been binned 2$\times$2, giving an effective scale of 
0.43\arcsec~pixel$^{-1}$. The solid line marked ``R"
indicates the orientation of the radio axis.
The extended emission regions are highly polarized, 
with PA perpendicular to the radius 
vectors, and the degree of polarization generally increasing with radius 
from the central nucleus. The radio axis (R) is not aligned with the 
UV/optical extensions in this object. \label{fig6}}
\end{figure}

\begin{figure}
\epsscale{0.9}
\plotone{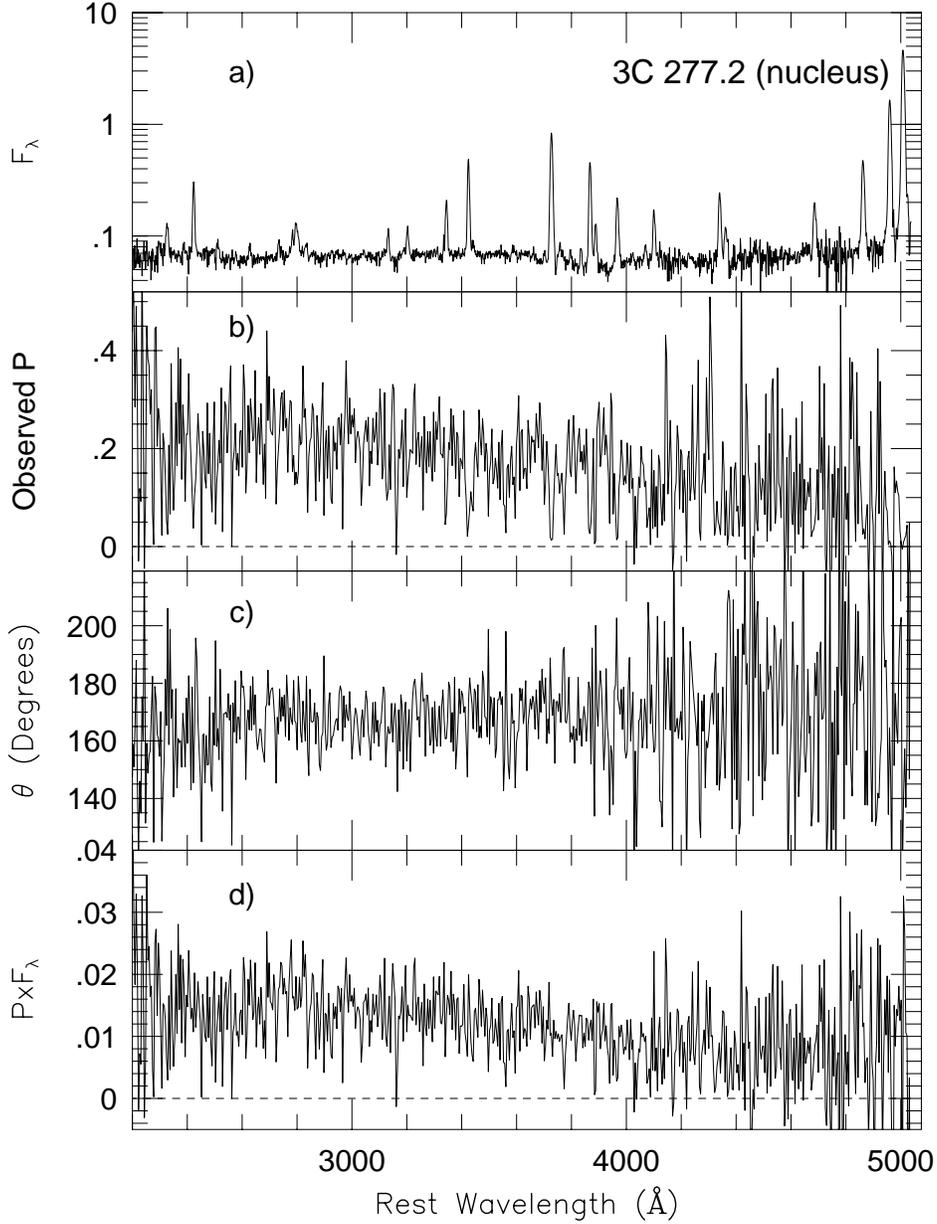}  
\caption{Spectropolarimetry of 3C 277.2 nucleus. 
From top to bottom are
the (a) total flux spectrum, \flam~plotted in logarithmic scale, 
(b) observed degree of polarization \p, (c) polarization position angle \pa,
and (d) polarized flux spectrum \pf. 
The flux scales are in units of 10$^{-16}$ ergs s$^{-1}$ cm$^{-2}$ \AA$^{-1}$.
All data except \flam~in (a) have been binned 3 pixels to improve S/N.
Broad Mg II line is evident in the total 
flux spectrum, although it is not as obvious in \pf. Stellar absorption
line features and the Balmer discontinuity around 3650 \AA~are also present. 
The narrow lines are unpolarized. \label{fig7}} 
\end{figure}

\begin{figure}
\plotone{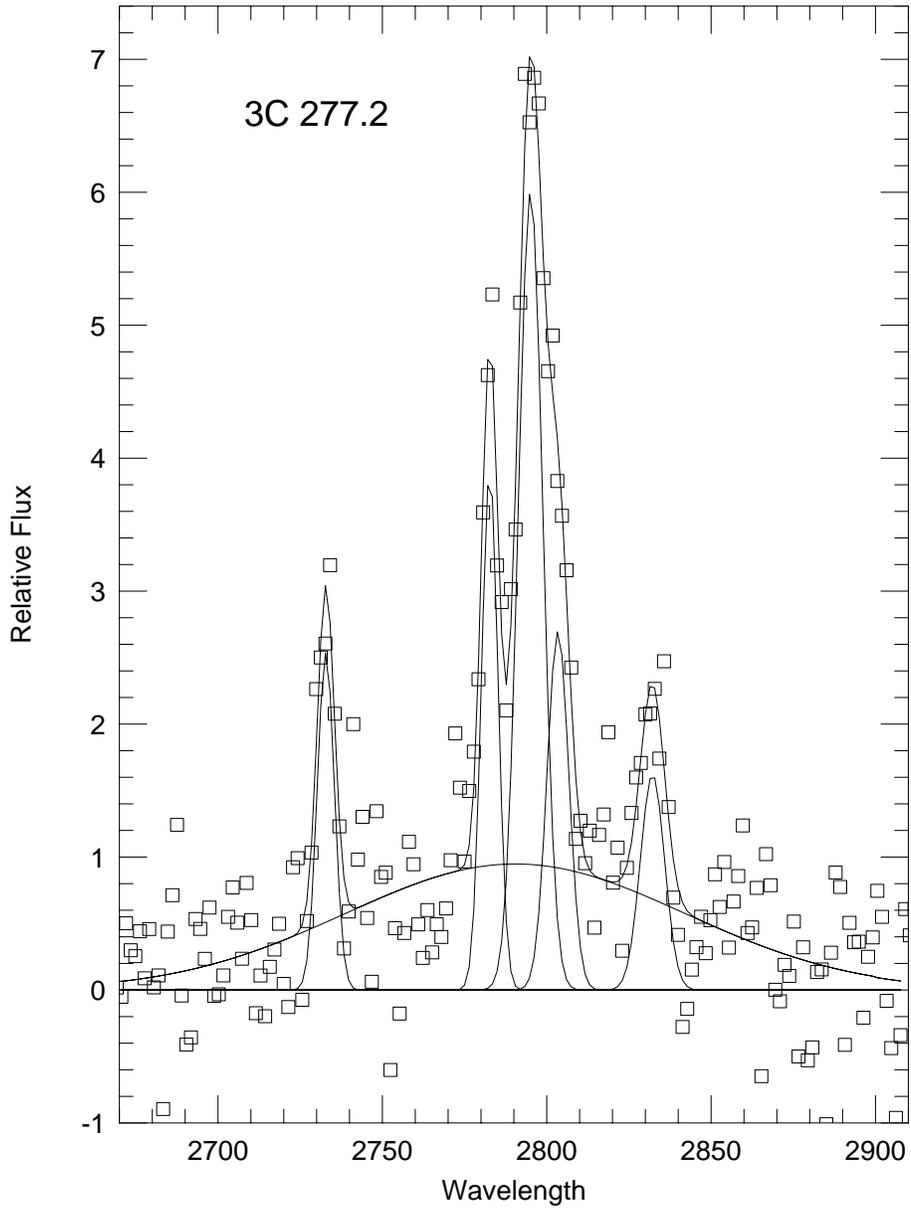}  
\caption{Fitting of Gaussian profiles in the total flux emission-line
complex of Mg II in 3C 277.2. The broad Mg II line has
FWHM = 13,000 $\pm$ 1500 \kms, similar to that of 3C 265. \label{fig8}}
\end{figure}

\begin{figure}
\epsscale{0.9}
\plotone{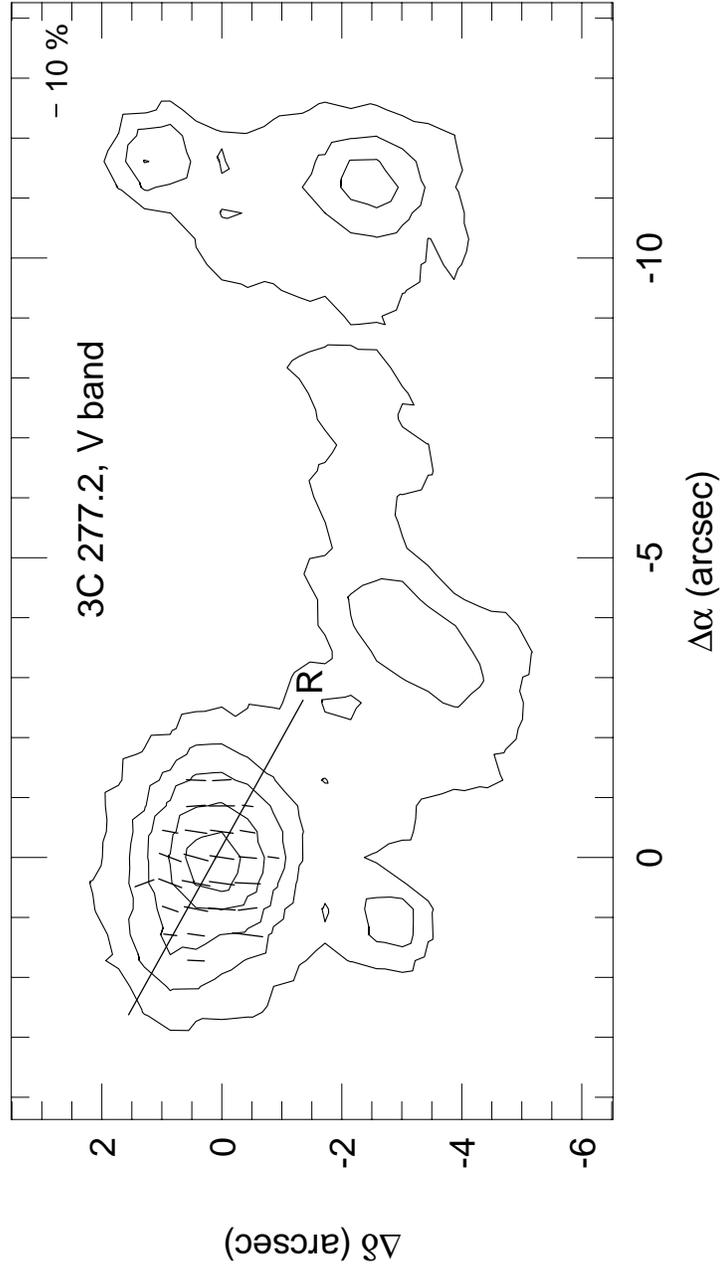}  
\caption{$V$-band imaging polarimetry of 3C 277.2. 
Contours are 2, 5, 10, 25, 60\% of peak intensity. Displayed as in Figure 6,
except the cutoff for vectors plotted is at 2.5$\sigma$~instead 
of 2$\sigma$.
High, significant polarization is detected near the nucleus. 
\p~is fairly constant throughout in the detectable regions and does not 
show the \p~gradient seen in 3C 265 and 3C 324. There are some extended 
emission regions to the west, whose polarization is below the detection 
limit of this image (\p~$<$ 4\%, 2$\sigma$). \label{fig9}}
\end{figure}

\begin{figure}
\plotone{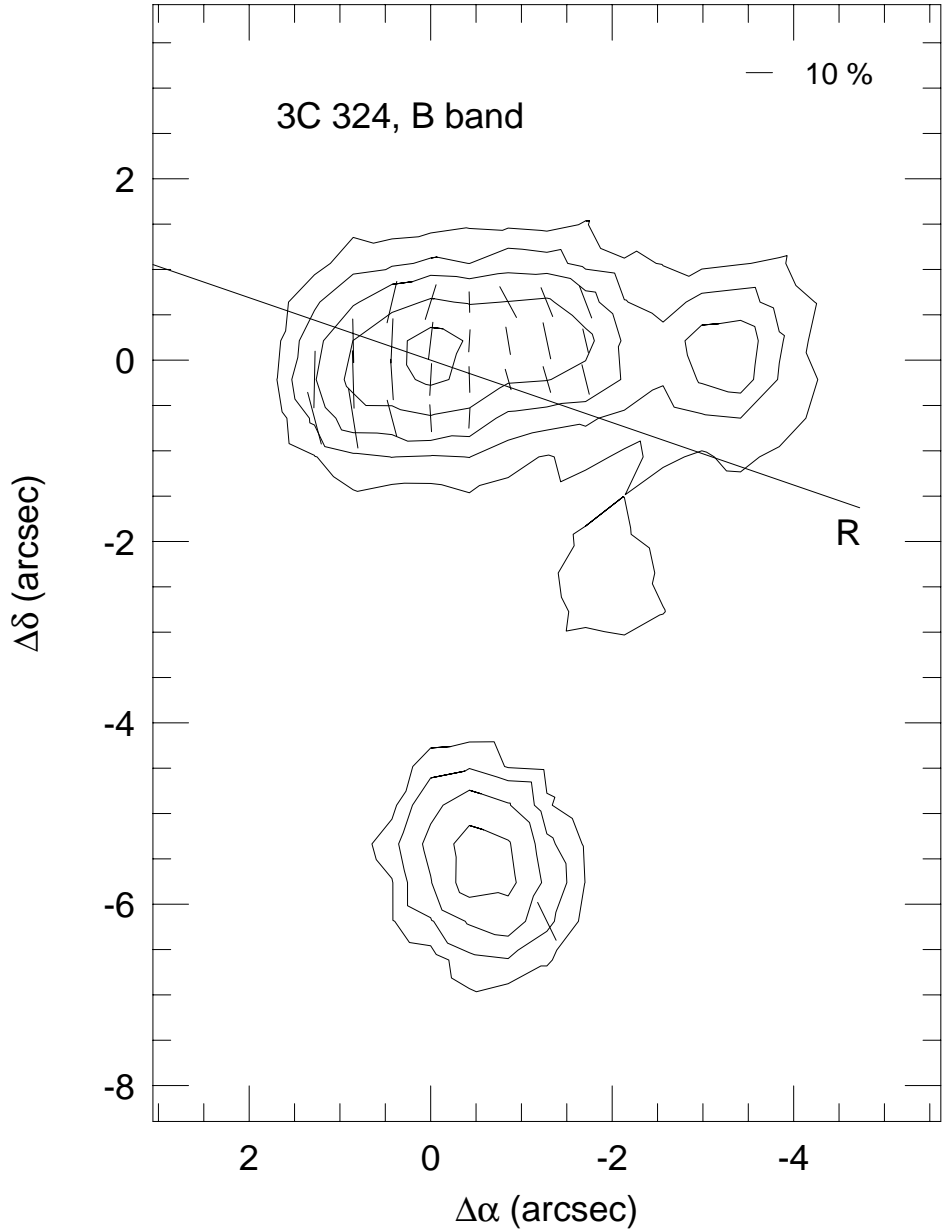}  
\caption{$B$-band imaging polarimetry of 3C 324, displayed as in Figure 6. 
Contours are 10, 17, 29, 49, 83.5\% of peak intensity.
\p~shows a slight increase from the nucleus to the eastern, 
but not the western extension. 
The component $\sim$ 3\arcsec~to the west of the nucleus is essentially 
unpolarized, suggesting that it is an independent galaxy dominated by 
starlight. \label{fig10}} 
\end{figure}

\begin{figure}
\epsscale{0.9}
\plotone{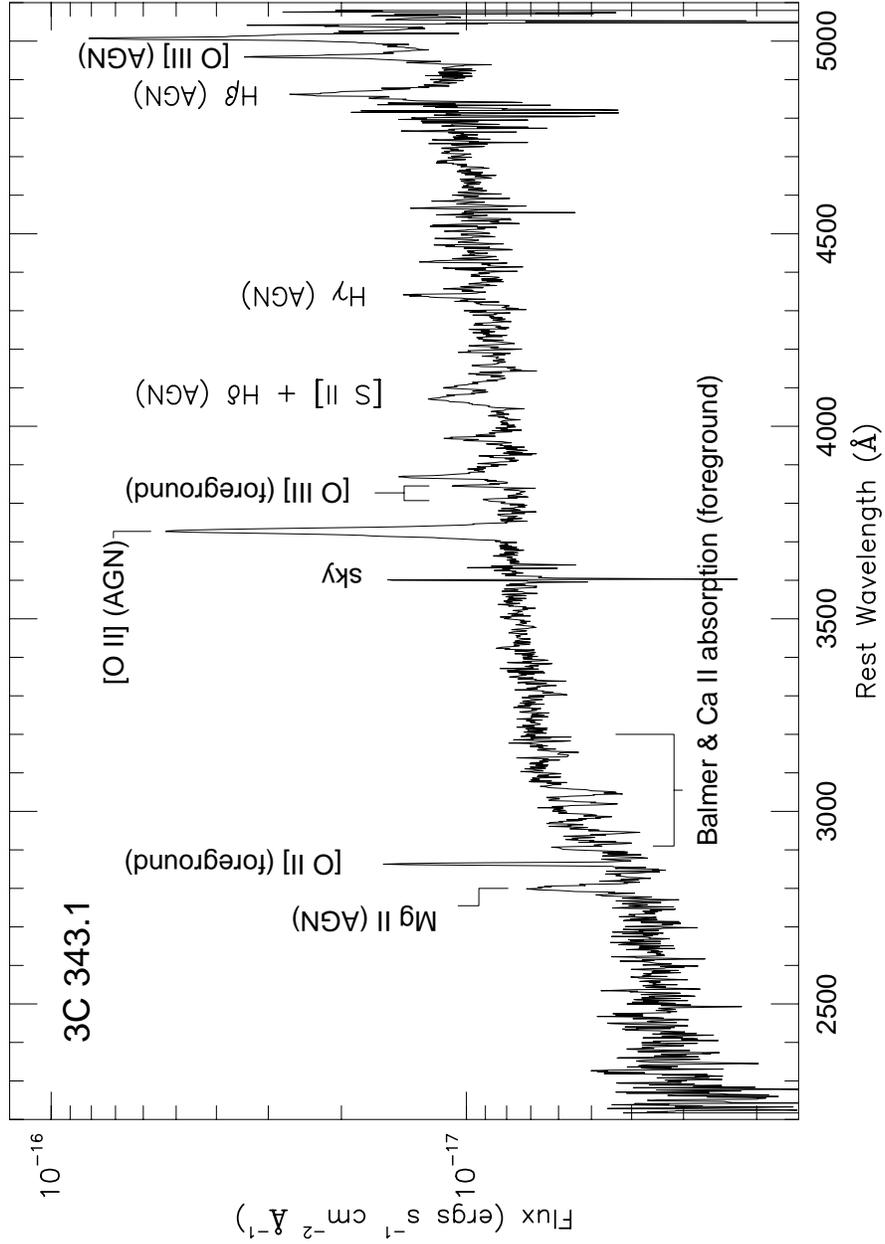}  
\caption{Observed total flux spectrum of 3C 343.1 nucleus, plotted
in logarithmic scale to show the faint features, and corrected for the
redshift of the AGN. Note the presence of strong 
high-$n$ Balmer absorption lines and broad wings in \hb, H$\delta$, and
Mg II, as well as the lack of high ionization [Ne V] and He II lines. 
The absorption-line redshift is substantially lower ($z=0.344$) than
the emission-line redshift of the AGN nucleus ($z=0.750$).
There are also emission lines ([O II] and \oiii) at the same redshift as
the absorption-line system. \label{fig11}}
\end{figure}

\begin{figure}
\epsscale{0.9}
\plotone{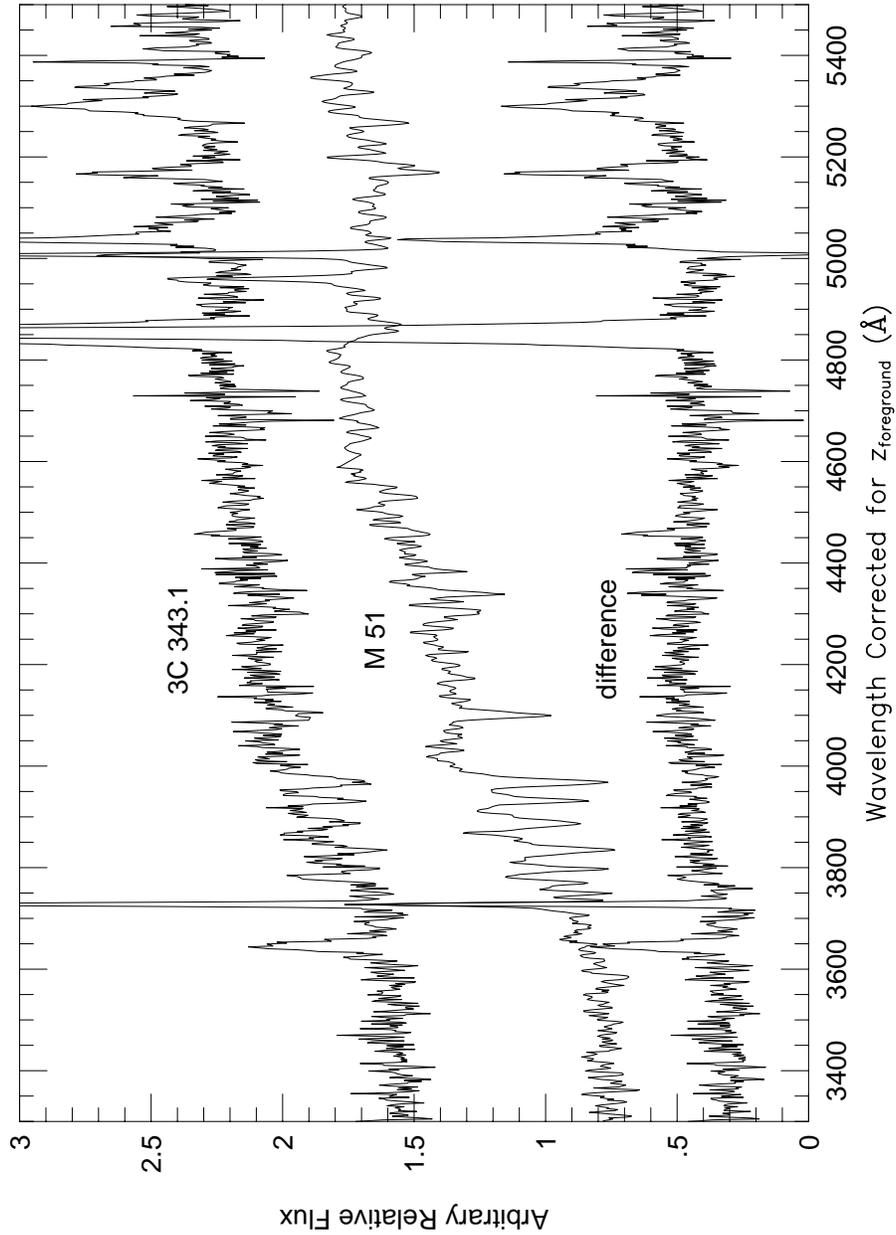}  
\caption{Decomposition of the nuclear spectrum of 3C 343.1 using
M51 as a template. {\it Top:} Observed spectrum of 3C 343.1; 
{\it Middle:} The spectrum of M51; {\it Bottom:} The difference spectrum. 
For clarity, all spectra have been arbitrarily shifted vertically, 
and corrected for the redshift of the absorption-line
system $z = 0.344$ belonging to the foreground emission-line galaxy. 
It has a spectrum similar to that of M51 and makes up 50--70\% of 
the observed flux toward 3C 343.1. \label{fig12}}
\end{figure}

\clearpage
% TABLE 1
\begin{deluxetable}{lcccc}
%\tiny
\tablewidth{0pc}
\tablecaption{Log of Polarimetric Observations}
\tablehead{
\colhead{Object} & \colhead{UT Date} & \colhead{Mode$^a$}  & \colhead{Exp. (s)}
& \colhead{Slit PA ($^\circ$)} }
\startdata
3C 265~~ & 1994 December 31 &  SPOL & 4 $\times$ 2310  & 142     \nl
         & 1995 January 28  &  SPOL & 4 $\times$ 2400  & 134     \nl
         & 1995 June 01     &  IPOL & 4 $\times$ 1620  & \nodata \nl
3C 277.2 & 1996 April 17    &  SPOL & 4 $\times$ 2400  & ~90     \nl
         & 1996 April 16    &  IPOL & 4 $\times$ 1260  & \nodata \nl
3C 324~~ & 1995 July 28     &  IPOL & 8 $\times$ ~900  & \nodata \nl
3C 343.1 & 1995 July 29     &  SPOL & 4 $\times$ 1800  & ~90     \nl
\tablenotetext{a}{SPOL = spectropolarimetry; IPOL = imaging polarimetry.}

\enddata
\end{deluxetable}

\clearpage

%TABLE 2
\begin{deluxetable}{lllllllll}
%\footnotesize
%\tiny
\tablecolumns{9}
\tablewidth{0pc}
\tablecaption{Observed Line Flux Ratios and Equivalent Widths}
\tablehead{
\colhead{} & \multicolumn{2}{c}{3C 265} & \multicolumn{1}{c}{} & \multicolumn{2}{c}{3C 277.2} & \multicolumn{1}{c}{} & \multicolumn{2}{c}{3C 343.1$^a$} \\
\cline{2-3} \cline{5-6} \cline{8-9} \\
\colhead{Line} & 
\colhead{Flux Ratio$^b$} & \colhead{EW} & \colhead{} &
\colhead{Flux Ratio$^b$} & \colhead{EW} & \colhead{} &
\colhead{Flux Ratio$^b$} & \colhead{EW} 
}
\startdata
C III \wave 2297  &  0.0389 & 3.10 && \nodata & \nodata && \nodata & \nodata \nl
C II] \wave 2326  &  0.233  & 19.5 && 0.1345 & 16.1 && \nodata & \nodata \nl
He II \wave 2386  &  0.0458 & 3.78 && 0.0327 & 4.01 && \nodata & \nodata  \nl
[Ne IV] \wave 2424 & 0.560 & 46.34 && 0.3905 & 44.15 && \nodata & \nodata  \nl
[O II] \wave 2470 & 0.0624 & 5.28 && \nodata & \nodata && \nodata & \nodata \nl
He II \wave 2512 + [Mg VII] \wave 2509 & 0.0560 & 5.24 && 0.0608 & 7.96 && \nodata & \nodata \nl
[Na VI] \wave 2569 ? & 0.0533 & 4.85 && \nodata & \nodata && \nodata & \nodata \nl
[Mg VII] \wave 2629 &  0.0654 &  6.00 && 0.0438 & 5.85 && \nodata & \nodata \nl
He II \wave 2733  & 0.0634 & 5.38 && 0.0637 & 8.34 && \nodata & \nodata\nl
[Mg V] \wave 2783 & 0.0721 & 5.47 && 0.0617 & 7.00 && \nodata & \nodata\nl
Mg II (b+n) \waves 2796, 2805 & 0.774 & 74.0 && 0.311 & 41.1 && 0.179 & 56.8 \nl
He I 2830 + O III \wave 2836 & 0.130 & 11.0 && 0.0601 & 7.73 && \nodata & \nodata \nl
[Ar IV] \wave 2854 & 0.0185 & 1.65 && \nodata & \nodata && \nodata & \nodata \nl
[Ar IV] \wave 2868 & 0.0147 & 1.36 && \nodata & \nodata && \nodata & \nodata \nl
[Mg V] \wave 2928  & 0.0240 & 2.27 && \nodata & \nodata && \nodata & \nodata \nl
He I \wave 2946    & 0.0208 & 1.96 && \nodata & \nodata && \nodata & \nodata \nl
O III \wave 3047   & 0.0264 & 2.65 && 0.023: & 2.8: && \nodata & \nodata  \nl
O III \wave 3133   & 0.1385 & 14.05 && 0.0842 & 10.30 && \nodata & \nodata  \nl
He I \wave 3188 + He II \wave 3203 & 0.177 & 18.0 & & 0.118 & 14.22 && \nodata
& \nodata  \nl
He I \wave 3188  & 0.0442 &  4.46 && \nodata & \nodata  && \nodata & \nodata\nl
He II \wave 3203 & 0.134  & 13.5 && \nodata & \nodata  && \nodata & \nodata\nl
[NeV] \wave 3346 & 0.3105 & 32.0 && 0.2814 & 33.43 && \nodata & \nodata  \nl
[NeV] \wave 3426 + O III \wave 3444 & 0.883 & 93.5 & & 0.778 & 92.3 && 0.0342 & 7.28 \nl
[Fe VII] \wave 3588 & 0.0585 & 6.53 && 0.011: & 1.3: && \nodata & \nodata \nl
[O II] \wave 3727 & 1.801 & 230.0 && 1.676 & 235.2 &&  2.031 & 432.4 \nl
H12 + H11 \waves 3750, 3771 &  0.143  & 18.94 && 0.0916 & 13.3 && \nodata & \nodata  \nl
H10 \wave 3798 &  0.0532 &  7.35 && 0.0430 & 6.41 && \nodata & \nodata  \nl
H9  \wave 3835 &  0.0559 &  8.02 && 0.0359 & 5.47 && \nodata & \nodata  \nl
[Ne III] \wave 3869 + H8 + He I & 1.130 & 159.17 &&   &  && 0.413 & 93.8 \nl
[Ne III] \wave 3869 + He I \wave 3868 & 0.910 & 128.8 && 0.817 & 125.5 && & \nl
H8 + He I \wave 3889 & 0.248 & 35.3 & & 0.123 & 19.0  && & \nl
Ca II K absorption &$-$0.0517 &$-$7.4 && \nodata & \nodata && \nodata & \nodata \nl
H$\epsilon$ + [Ne III] \wave 3967 & 0.387 & 55.12 && 0.360 & 52.7 && 0.203 & 42.5 \nl
[S II] \wave 4071 & 0.0713 & 10.24 && 0.0412 & 5.78 && 0.272 & 51.1  \nl
H$\delta$           &  0.287  & 41.76 && 0.229  & 31.8 && 0.114 & 21.5  \nl
[Fe V] \wave 4227 &  0.0787 & 12.30 && 0.029: & 4.0: && \nodata & \nodata  \nl
H$\gamma$      &  0.4722  & 71.7 && 0.3445 & 48.7 && \nodata & \nodata  \nl
[O III] \wave 4363  & 0.2018  & 30.40 && 0.157 & 21.9 && \nodata & \nodata  \nl
He I \wave 4471    & 0.0347 & 5.01 && \nodata & \nodata && \nodata & \nodata \nl
He II \wave 4686    & 0.273  & 37.0 && 0.293 & 35.5 && \nodata & \nodata  \nl
[Ar IV] \wave 4712 & 0.0667 & 9.06 && \nodata & \nodata && \nodata & \nodata \nl
[Ar IV] \wave 4740 & 0.0543 & 7.36 && \nodata & \nodata && \nodata & \nodata \nl
H$\beta$            & 1.000  & 121.59 && 1.000 & 110.0 && 1.000 & 117.7 \nl
[O III] \wave 4959  & \nodata & \nodata &&  3.856 & 370.6 && 0.746 & 75.3 \nl 
[O III] \wave 5007  & \nodata & \nodata && 11.67  & 995.5 && 2.21 & 200.0 \nl        
\tablenotetext{a}{Flux ratio and EWs are measured from galaxy-subtracted
spectrum.} 
\tablenotetext{b}{Flux ratio relative to H$\beta$; colon denotes uncertainty $\simgt$ 20\%.}

\enddata

\end{deluxetable}

\end{document}